\documentclass{article}
\usepackage{jheppub}
\usepackage{tensor}
\usepackage{multirow}
\pdfoutput=1

\newcommand{\roughly}[1]{\mathrel{\raise.3ex\hbox{$#1$\kern-0.85em\lower1ex\hbox{$\sim$}}}}

\newcommand{\lsim}{\roughly<}
\newcommand{\gsim}{\roughly>}
\def\pd{\partial}
\newcommand{\abs}[1]{\left|{#1}\right|}

\def\cG{{\cal G}}
\def\cH{{\cal H}}

\def\cR{{\cal R}}

\def\nn{\nonumber}
\def\({\left(}
\def\){\right)}
\def\[{\left[}
\def\]{\right]}

\newbox\charbox
\newbox\slabox
\def\slsh#1{{      
        \setbox\charbox=\hbox{$#1$}
        \setbox\slabox=\hbox{$/$}
        \dimen\charbox=\ht\slabox
        \advance\dimen\charbox by -\dp\slabox
        \advance\dimen\charbox by -\ht\charbox
        \advance\dimen\charbox by \dp\charbox
        \divide\dimen\charbox by 2
        \raise-\dimen\charbox\hbox to \wd\charbox{\hss/\hss}
        \llap{$#1$}
}}

\def\exd{{\hbox{d}}}

\newcommand{\f}{\frac}

\def\nn{\nonumber}

\def\MP{M_{p}}
\def\bnow{b_\star}

\def\kk{{\rm{KK}}}

\def\bea{\begin{eqnarray}}
\def\eea{\end{eqnarray}}
\def\be{\begin{equation}}
\def\ee{\end{equation}}

\def\ssM{{\scriptscriptstyle M}}
\def\ssN{{\scriptscriptstyle N}}
\def\ssP{{\scriptscriptstyle P}}

\def\pref#1{(\ref{#1})}

\def\bal#1\eal{\begin{align}#1\end{align}}
\def\ba*#1\ea*{\begin{align*}#1\end{align*}}

\title{Goldilocks Models of Higher-Dimensional Inflation (including modulus stabilization)}

\author[a,b]{C.P.~Burgess,}
\author[a,b]{Jared J.H.~Enns,}
\author[a,b]{Peter Hayman}
\author[c]{and Subodh P.~Patil}
\affiliation[a]{Physics \& Astronomy, McMaster University, Hamilton, ON, Canada, L8S 4M1}
\affiliation[b]{Perimeter Institute for Theoretical Physics, Waterloo, ON, Canada, N2L 2Y5}
\affiliation[c]{Department of Theoretical Physics, University of Geneva, 24 Quai Ansermet, Geneva, CH-1211, Switzerland}

\emailAdd{cburgess@perimeterinstitute.ca, ennsjj@mcmaster.ca, haymanpf@mcmaster.ca, subodh.patil@unige.ch}

\date{\today}

\abstract {We explore the mechanics of inflation within simplified extra-dimensional models involving an inflaton interacting with the Einstein-Maxwell system in two extra dimensions. The models are Goldilocks-like inasmuch as they are just complicated enough to include a mechanism to stabilize the extra-dimensional size (or modulus), yet simple enough to solve explicitly the full extra-dimensional field equations using only simple tools. The solutions are {\em not} restricted to the effective 4D regime with $H \ll m_\kk$ (the latter referring to the characteristic mass splitting of the 
Kaluza-Klein excitations) because the full extra-dimensional Einstein equations are solved. This allows an exploration of inflationary physics in a controlled calculational regime away from the usual four-dimensional lamp-post. The inclusion of modulus stabilization is important because experience with string models teaches that this is usually what makes models fail: stabilization energies easily dominate the shallow potentials required by slow roll and so open up directions to evolve that are steeper than those of the putative inflationary direction. We explore (numerically and analytically) three representative kinds of inflationary scenarios within this simple setup. In one the radion is trapped in an inflaton-dependent local minimum whose non-zero energy drives inflation. Inflation ends as this energy relaxes to zero when the inflaton finds its own minimum. The other two involve power-law scaling solutions during inflation. One of these is a dynamical attractor whose features are relatively insensitive to initial conditions but whose slow-roll parameters cannot be arbitrarily small; the other is not an attractor but can roll much more slowly, until eventually transitioning to the attractor. The scaling solutions can satisfy $H > m_\kk$, but when they do standard 4D fluctuation calculations need not apply. When in a 4D regime the solutions predict $\eta \simeq 0$ and so $r \simeq 0.11$ when $n_s \simeq 0.96$ and so are ruled out if tensor modes remain unseen. Assessment of general parameters is difficult until a full 6D fluctuation calculation is done.}

\begin{document}
\maketitle
\section{Introduction and summary}

The observation of primordial fluctuations in the Cosmic Microwave Background (and in the pattern of large-scale structure) provides a rare observational window into physics at energies much higher than can at present be probed in any other way. But for now we see through this glass, darkly: the glimpse we are provided informs (but does not dictate) what the new high-energy physics ultimately is.

Two features are shared by all glimpses so far. The first is the need for a sufficiently long period of accelerated expansion (usually inflationary acceleration \cite{Guth:1980zm,Linde:1981mu,Albrecht:1982wi} of an earlier universal expansion or the acceleration that reverses an earlier period of universal contraction \cite{Gasperini:1992em,Khoury:2001wf,Ashtekar:2007em,Buchbinder:2007ad,Easson:2011zy,Kounnas:2011gz,Brandenberger:2013zea} --- for some reviews with references see \cite{Baumann:2009ds,Novello:2008ra,Brandenberger:2016vhg}), whose role is partly to clean the cosmic canvas of all earlier patterns. The second is the generation of the observed inhomogeneities by amplifying initially small vacuum fluctuations so that they can be writ large across the sky \cite{Mukhanov:1981xt,Linde:1982uu,Hawking:1982cz,Guth:1982ec,Bardeen:1983qw}. It is remarkable that such a simple source describes the pattern of observed fluctuations so well \cite{Ade:2015lrj}.

To date, inflationary models have received the most attention, both because of their successes describing the observations and because they (at least at present) provide the best-controlled framework within which the implications of the very early universe can be crisply explored. Yet, existing models often seem unconvincing. Potential problems can include extreme sensitivity to initial conditions, the requirement of unnaturally flat scalar potentials, the dangers of eternal inflation and difficulties ending inflation everywhere in the universe, relatively poor understanding of reheating, and so on \cite{Brandenberger:1999sw,Burgess:2011fa,Ijjas:2015hcc}. 

Many of these problems have their root in the UV-completion within which the inflationary evolution occurs. For instance, initial conditions refer to earlier epochs involving still higher energies. Shallow potentials are unnatural because of loops involving very massive particles. Eternal inflation is common in a landscape, but the existence and features of a landscape depend on the number of fields present at very high energies. Such problems cannot be addressed without specifying the broader context from which the inflaton field arises.

The goal of this paper is to develop particularly simple models of inflation in which new types of UV questions can be reliably asked. A central feature of the models is that they are explicitly extra-dimensional, which is done since extra dimensions are very plausible parts of any UV completion at the energies of likely inflationary interest, particularly if string theory plays a role in the unitarization of gravity. We also do not restrict to a single extra dimension, and consequently explore outside the relatively well-explored cosmology \cite{Cline:1999yq,Flanagan:1999cu,Kim:2000hi,Lesgourgues:2000tj} of the Randall-Sundrum \cite{Randall:1999ee} sandbox. Although there are extensive studies of inflation with a stringy provenance \cite{Linde:2005dd,Burgess:2007pz,Baumann:2009ni,Cicoli:2011zz} (which also involve more than one extra dimension, and can do well when compared with observations \cite{Burgess:2013sla,Silverstein:2015mll}), comparatively few of these studies actually involve explicit solutions of the extra-dimensional field equations (see however \cite{vanNierop:2011di}). Most studies instead identify scalar fields within a 4D effective theory below the Kaluza-Klein (KK) scale,\footnote{Here and throughout we use hats to distinguish 6D quantities from 4D quantities where ambiguities may arise. By $m_\kk$, we are refer to the characteristic mass splitting of the KK resonances.} $m_\kk$. The existence of inflation is then inferred using only the 4D effective Einstein equations. While this is a legitimate procedure, it usually requires all derivatives (including in particular the 4D Hubble scale, $H$) to be smaller than $m_\kk$, and so forces inflationary searches to proceed only underneath an explicitly four-dimensional lamppost. 

An important point about the models we explore here is that all extra-dimensional field equations are solved explicitly and so there is no requirement to be within the domain of validity of effective 4D methods. In particular, we identify several examples with $H$ greater than the mass splitting $m_\kk$, which would not be possible to explore at all if only 4D effective methods were available. In the strict de Sitter limit, the fact that the isometry group contains special conformal transformations requires all KK excitations with spin $s > 3/2$ to posses a mass gap $\Delta m^2_{\rm KK} > s(s-1)H^2$ (through unitarity of the representation \cite{Arkani-Hamed:2015bza}; in the case of spin two the former is more widely known as the Higuchi bound \cite{Higuchi:1986py}). However we note that quasi de-Sitter backgrounds such as those of the `extended-inflation' (or power law) variety \cite{Abbott:1984fp,Yokoyama:1987an,Liddle:1988tb,PhysRevD.32.1316,La:1989za,Weinberg:1989mp,Liddle:1992wi,Copeland:1997et} break conformal symmetry, and it is not clear an analogous bound can be derived on such backgrounds\footnote{In any case, all KK excitations above any mass gap would have their mass splittings $m_\kk$ determined by the inverse radius of compactification, and it is with respect to this scale that we find backgrounds with $H > m_\kk$.}.

Power-law inflation is also interesting for another reason: it robustly predicts $\epsilon = - \dot{H}/ H^2$ to be constant and so the second slow-roll parameter is $\eta := \dot{\epsilon}/( H \epsilon) = 0$. When power-law evolution occurs with $H \ll  m_\kk$, standard 4D calculations of primordial fluctuations can apply and then the prediction $\eta = 0$ implies the scalar-to-tensor ratio, $r$, and scalar spectral index, $n_s$, are related by \cite{Liddle:1992wi}
\be
   r = \frac83 \, (1 - n_s)  \,,
\ee
which implies $r \simeq 0.11$ if $n_s \simeq 0.96$. This is already in some tension with the latest 95\% c.l.~upper limits, $r < 0.07$ \cite{Array:2015xqh} and so these models may soon be excluded (unless primordial tensor modes show up very soon). Of course, predictions are less clear when $H > m_\kk$ since, in this regime, standard 4D calculations need not capture the right fluctuation spectrum.

A second key feature of the models examined here is that they explicitly provide a potential that can stabilize all of the moduli that are generically present within extra dimensions. Indeed, experience with modulus stabilization in Type IIB string models \cite{Freund:1980xh,Giddings:2001yu,Dasgupta:1999ss,Kachru:2003aw} and its influence on inflationary models \cite{Kachru:2003sx,BlancoPillado:2004ns} shows that modulus stabilization is usually crucial to any understanding of any extra-dimensional inflationary dynamics. It is crucial because the very shallow potentials usually required by inflation are easily swamped by the physics of modulus stabilization, which generically provides a steeper, non-inflating, direction in field space along which the system prefers to evolve. Although extra-dimensional inflationary models have been explored extensively since the early 1980s \cite{Freund:1982pg,RandjbarDaemi:1983jz,Shafi:1984ha,Abbott:1984ba,Sahdev:1988fp} (usually with compact extra dimensions, though see also \cite{Kaloper:2000jb}), most (with a few exceptions \cite{Okada:1984cv,Maeda:1984gq,WETTERICH1985309,Gunther:2000jj,Gunther:2000yb,vanNierop:2011di,BlancoPillado:2011me}) ignored modulus stabilization usually due to the lack of tools with which to study it. But since the development of tools like flux-stabilization \cite{Giddings:2001yu, Grana:2005jc}, the state of the art for any modern understanding requires a treatment of modulus stabilization. 

Finally, our models also enjoy a third important feature: they are {\em Goldilocks} models in the sense that they are just complicated enough to allow the appearance of many of the features (like flux stabilization) that also arise in more involved stringy models, yet they are simple enough not to require overly fancy tools (usually more to do with the complications of string physics than about the mechanics of inflation) beyond those already in the toolbox of most cosmologists. Because of this, we hope these models will be used in future work to seek higher-dimensional signatures in the predictions for primordial fluctuations. This balance between new phenomena and ease of calculation is achieved by working with a minimal number of dimensions,\footnote{Yet with more than the single extra dimension of RS models, since one dimension often has very special features that do not extend to higher dimensions.} with only the minimal fields required for inflation and modulus stabilization. Indeed we work here with six-dimensional models because these are among the simplest to stabilize, and are among the earliest for which modulus stabilization was systematically developed \cite{RandjbarDaemi:1982hi,Salam:1984cj,Maeda:1985bq} (for similar reasons, six-dimensional models have been used as toy laboratories for extra-dimensional tunneling problems \cite{BlancoPillado:2009di,BlancoPillado:2009mi,Brown:2010bc,Brown:2010mf,Brown:2010mg,Brown:2011gt}). We verify our solutions remain deep within the regime of validity of semiclassical methods \cite{Burgess:2003jk,Burgess:2009ea}, and as a result avoid models whose success relies on UV-sensitive assumptions such as higher-derivative terms.

We examine in particular an Einstein-Maxwell-Scalar system in six dimensions, within which the Maxwell field provides the required flux stabilization and the scalar, $\phi$, is essentially a higher-dimensional inflaton (but is {\em not} restricted to be the dilaton, such as often arises in 6D supergravity models). Most of the interesting dynamics arise due to the interplay between the evolution of the inflaton and of the extra-dimensional modulus. Since we use spherical extra dimensions, there is only a single modulus: the radion, $\psi$. Because both the inflaton and radion evolve as the universe expands, relations become possible between the number of inflationary $e$-foldings and the change in size of the extra dimensions. In one case, we find these are related by $N_e = 2 \ln(b_f/b_0)$, where $b$ is the physical radius of the extra dimensions. This gives the tantalizing numerology that $N_e \simeq 60$ if $b_f/b_0 \sim 10^{13}$ or  $N_e \simeq 70$ if $b_f/b_0 \sim 10^{15}$ (as could be the case if $b_0$ were near the Planck scale and $b_f$ near the weak scale, for instance).

We find three different inflationary regimes within this model, two of which involve power-law inflation and allow the regime $H > m_\kk$. They have the following properties:

\medskip\noindent{\em Cradle Regime:} 

\medskip\noindent This corresponds to a situation where the radion gets trapped early by a local minimum of its potential, before the inflaton has yet found its own minimum. Although the potential is tuned to vanish when both fields are minimized, it does not vanish in this case while the inflaton is still rolling and so inflation is driven by the energy at the radion's local minimum. As the inflaton rolls, the radion gets slowly lowered towards zero, at which point inflation ends. 

This regime is both unusual and commonplace in extra-dimensional models. It is commonplace in the sense that the radion does what stabilized moduli do once they settle into a local minimum during an inflationary regime. However, such moduli are usually integrated out when inflation is analyzed within an effective 4D theory, so what is unusual is our ability to track the approach to the minimum and its evolution explicitly. We can do so because we can follow all of the extra-dimensional dynamics explicitly.

This regime tends to satisfy $H \lsim m_\kk$ and so standard 4D calculations of primordial fluctuations can apply (where one can generically expect features and suppression of power at long wavelengths \cite{Chluba:2015bqa}). If they do we find slow-roll parameters governed by the slow inflaton roll, which resembles chaotic inflation \cite{Linde:1983gd} because the inflaton is high up the side of a potential hill. In the model described here in detail, we find a range $0.13 \lsim r \lsim 0.18$ corresponding to $70 \gsim N_e \gsim 50$ $e$-foldings, which is higher than the best current bounds. Athough we have not systematically explored for lower values in this model, smaller $r$ goes with larger $N_e$ (with $r \sim 0.1$ if we allow $N_e \simeq 90$) .

\medskip\noindent{\em Attractor Power-law Regime:} 

\medskip\noindent This regime corresponds to a power-law scaling solution that arises when the inflaton's 6D potential dominates the energy density and is approximately exponential. Although other scaling solutions can also arise in other parts of parameter space (such as the example described in the next item), this one is special inasmuch as it is an {\em attractor} (in the original dynamical sense of the word, wherein it is the endpoint of a broad class of initial conditions). As a result, is often shows up when the field equations are integrated numerically. Inflationary attractors of this type have arise elsewhere within a 4D context \cite{Maeda:1987xf,Muller:1989rp,GarciaBellido:1995kc,Copeland:1997et,Ferreira:1997hj}.

This solution enjoys several noteworthy features, including that it can naturally end (because the potential is no longer approximately exponential) once the radion and inflaton find their minima. There is generally an overshoot problem with the radion, however, since its minimum is not very deep, though one can tweak the model to avoid this. This solution also generically allows $ H /  m_\kk > 1$ and moreover this ratio can grow with time. The main drawbacks of this solution for inflationary phenomenology are its prediction of a lower bound $\epsilon \ge \frac12$ and, if it is to end in a regime with a minimum for the radion, its prediction of an upper bound $N_e \lesssim 35$ for the number of $e$-foldings. Whether the lower bound on $ \epsilon$ is a problem depends on the method of generating primordial fluctuations, and whether standard results survive in the cases where $ H >  m_\kk$. The upper limit on $N_e$ is more robust but might be acceptable if the attractor solution arises as the endpoint of a different inflationary solution (such as the one described next) for which $ \epsilon$ can be smaller.

\medskip\noindent{\em Slow-Roll Power-law Regime:} 

\medskip\noindent This regime corresponds to one of the classes of scaling solutions that are not attractors. In particular, the solution of interest here arises when the inflaton potential and the extra-dimensional curvature have similar size and together dominate the energy density. This solution is not an attractor because small perturbations in one direction in field-space grow with time (in numerical solutions often eventually crossing over to the above attractor solution). On one hand, this unstable direction provides a natural way of ending inflation, while on the other hand its existence means this solution requires specially arranged initial conditions. There is no lower bound in this solution for the slow-roll parameters $\epsilon$, nor is there an upper bound on the number of $e$-foldings, $N_e$. The Hubble and KK scales can satisfy $ H >  m_\kk$, however, for this solution, their ratio does not evolve in time. A phenomenological disadvantage of this model is that the conditions for its existence preclude the possibility of it ending in a regime for which the radion has a viable minimum for $V$, so a realistic scenario must invoke another mechanism for modulus stabilization beyond the simplest flux-stabilization mechanism. 

\bigskip\noindent
The remainder of the paper is organized as follows. Section \ref{section:Systems} lays out the action and field equations of our Goldilocks system. After identifying the full set of field equations in six dimensions, section \ref{subsection:6D} shows that their simplification using a homogeneous symmetry {\it ansatz} leads to a reduced set of equations and shows these to be completely equivalent to the homogeneous field equations of a particular four-dimensional two-scalar field theory. The 4D system provides a consistent truncation \cite{DUFF1985355} of the full system, as is known to occur in 6D reductions on a sphere \cite{Gibbons:2003gp}. (We verify the consistency of this truncation in Appendix \ref{App:Consistency}.) Because the truncation is consistent, any solution to the 4D system can be dimensionally oxidized to a solution of the full 6D system. In section \ref{subsec:Analytic}, we quantify the conditions required to stabilize the radion with flux, and then derive the analytic power-law solutions that often dominate numerically. Section \ref{section:numerics} describes explicit numerical evolution for the three illustrative scenarios described above. 

\section{The system}
\label{section:Systems}

This section sets out the action and field equations to be solved for the solutions of interest in later sections.

\subsection{Action and field equations}
\label{subsection:6D}

The system of interest is a 6D Einstein-Maxwell-Scalar field theory, with action 
\bea
 && \qquad\qquad S = - \int \exd^6 x \, \sqrt{-g_{(6)}} \;  \Bigl(L_{\rm grav} + L_{\rm mat} \Bigr) \qquad \hbox{with}\nn\\
&&L_{\rm grav} = \frac{1}{2\kappa^2} \, \cR \qquad \hbox{and} \qquad
L_{\rm mat} =  \f 1 4 F_{\ssM\ssN}F^{\ssM\ssN} +  \f 1 2 \, \pd_\ssM\phi \,\pd^\ssM \phi +V(\phi) \,,
\eea
where $F_{\ssM\ssN} = \partial_\ssM A_\ssN - \partial_\ssN A_\ssM$ is an abelian gauge field strength while $g_{(6)}$ and $\cR = g^{\ssM\ssN} \cR_{\ssM \ssN}$ are the determinant and Ricci scalar for the 6D metric,\footnote{Our metric is `mostly plus' and we adopt Weinberg's curvature conventions \cite{Weinberg:1972kfs}.}  $g_{\ssM \ssN}$. Where explicitness is necessary, we choose a scalar potential of the form
\be \label{6Dpot}
 V(\phi) = V_0 \Bigl( e^{-\beta_1 \phi} - e^{-\beta_2 \phi} \Bigr) + \Lambda \,,
\ee
which (for $V_0$ and the $\beta_i$ positive) is minimized by 
\be
 \phi_\star = \frac{1}{\beta_2 - \beta_1} \, \ln \left( \frac{\beta_2}{\beta_1} \right) \,.
\ee
We may adjust $\Lambda$ to ensure that the 4D part of spacetime is flat once the extra-dimensional metric also satisfies its field equations.

The field equations following from this action are
\be
 \Box\, \phi  - V'(\phi) = 
 \nabla_\ssM F^{\ssM\ssN}  =
 \cG_{\ssM \ssN} + \kappa^2 T_{\ssM\ssN} = 0 \,, 
\ee
where $\cG_{\ssM\ssN} = \cR_{\ssM \ssN} - \frac12 \, \cR \, g_{\ssM\ssN}$ is the metric's Einstein tensor and the stress-energy is
\be
 T_{\ssM \ssN} = \partial_\ssM \phi \, \partial_\ssN \phi + F_{\ssM \ssP} {F^\ssP}_\ssN - g_{\ssM \ssN} L_{\rm mat} \,.
\ee

For cosmological applications, we seek time-dependent solutions where four dimensions have an FRW form (with, for simplicity, flat spatial slices) and the extra two dimensions have the geometry of a sphere:
\bal   \label{metric}
 \exd \hat s^2 = \hat g_{\mu\nu} \, \exd x^\mu \exd x^\nu + g_{mn} \, \exd y^m \exd y^n
 = -\exd \hat t^2 + \hat a^2(\hat t\, ) \, \delta_{ij} \,\exd x^i \, \exd x^j + b^2(\hat t \,) \gamma_{mn}(y) \, \exd y^m \exd y^n \,.
\eal
Here, $\gamma_{mn}$ is the standard round metric for the unit 2-sphere, and $\hat a$ and $b$ are the time-dependent scale factors for the three- and two-dimensional spatial dimensions, respectively, whose coordinates are denoted $x^i$ and $y^m$. The hats on $\hat g_{\mu\nu}$, $\hat a$ and $\hat t$ are used to distinguish them from the corresponding quantities defined in the 4D Einstein frame, defined below. We similarly take the scalar field to be homogeneous, $\phi = \phi(\hat t\,)$ and the only nonzero components of the Maxwell field to be $F_{mn} = f \, \epsilon_{mn}$, where $\epsilon_{mn}$ is the Levi-Civita tensor built from $g_{mn}$. 

With these {\em ans\"atze}, the homogeneous Klein-Gordon equation reads
\bal
  {\phi}'' + \Bigl( 3 \hat H + 2 \cH \Bigr) \phi' + \frac{\partial V}{\partial \phi} = 0\,, \label{6EOM3}
\eal
where $\hat H := {\hat a'}/\hat a$ and $\cH := b'/b$ and primes denote $\exd/\exd \hat t$ (we reserve over-dots for later use as $\exd/\exd t$ in 4D Einstein frame). The Maxwell equation and gauge-field Bianchi identity similarly become
\be \label{MaxwellBianchi}
 \partial_m f = 0 \quad \hbox{(Maxwell)} \qquad \hbox{and} \qquad
 ( f b^2 )' = 0 \quad \hbox{(Bianchi)} \,,
\ee
while quantization of the extra-dimensional magnetic flux implies
\be \label{quantiz}
\int_{S^2} F = 4\pi f b^2 = \frac{2\pi n}{e} 
\qquad \hbox{and so} \qquad
f = \frac{\mathfrak{f}}{b^2} \quad \hbox{with} \quad \mathfrak{f} := \frac{n}{2e} \,, 
\ee
with $n$ an integer and $e$ the Maxwell field's coupling constant. Clearly \pref{quantiz} satisfies both of eqs.~\pref{MaxwellBianchi}.

The stress-energy tensor for this metric {\it ansatz} becomes diagonal, with energy density, $\rho$, and 3D and 2D pressure, $p_{(3)}$ and $p_{(2)}$ given by
\bal
\rho &= \f 1 2 \left[ \left({\phi}' \right)^2 + f^2\right] + V \nn\\
p_{(3)} &= \f 1 2 \left[ \left( {\phi}' \right)^2 - f^2\right] - V \\
p_{(2)} &= \f 1 2 \left[ \left( {\phi}' \right)^2 + f^2\right] - V \,,\nn
\eal
and so the Einstein equations become
\bal
3\left(\f {{\hat a'}}{\hat a} \right)^2 + \left(\f {{b'}}{b} \right)^2 + 6\left(\f {{\hat a'}{b'}}{\hat ab}\right) + \f 1 {b^2} &= \kappa^2 \left\{  \f 1 2 \left[ \left({\phi}'\right)^2 + \f {\mathfrak{f}^2}{b^4} \right] + V \right\} \nn\\ 
2\left(\f {{\hat a''}}{\hat a}  + \f {{b''}}{b} \right) + \left( \f {{\hat a'}}{\hat a} \right)^2 + \left( \f {{b'}}{b}\right)^2 + 4\left(\f {{\hat a'} {b'}}{\hat ab}\right) +  \f 1 {b^2}&= \kappa^2 \left\{ \f 1 2 \left[ -\left({\phi}'\right)^2 + \f {\mathfrak{f}^2}{b^4} \right] + V \right\}  \label{EE} \\
 \f {{b''}}{b} + 3\left[ \f {{\hat a''}}{\hat a} + \left(\f {{\hat a'}}{\hat a} \right)^2 \right] + 3\left(\f {{\hat a'}{b'}}{\hat ab}\right) &= \kappa^2 \left\{ -\f 1 2 \left[ \left( {\phi}' \right)^2 + \f {\mathfrak{f}^2}{b^4} \right] + V \right\}. \nn
\eal

We have verified that solutions to these reduced equations also satisfy all of the other 6D field equations. They do so because our {\it ansatz} is the most general one consistent with translational and rotational invariance of the three spatial dimensions together with the maximal SO$(3)$ symmetry of the two extra dimensions. 

\subsubsection*{Domain of validity of approximations}

In later sections, we seek classical solutions to these equations and so must remain within the domain of validity of semiclassical methods. This requires curvatures are small in 6D Planck units, and so also requires restricting to configurations with correspondingly small stress energy 
\be
 \kappa \hat H^2 , \, \kappa \cH^2 , \, \kappa/b^2 \ll 1\qquad \hbox{and so} \qquad 
  \kappa^3 V_\star  \ll 1 \quad \hbox{and} \quad
  \kappa^3 F_{mn}F^{mn} \sim \frac{\kappa^3 \mathfrak{f}^2}{b^4} \ll 1 \,. \label{validity}
\ee

\subsection{Dimensional reduction}
\label{subsection:6D}

It is useful to rewrite the above 6D field equations in terms of an equivalent effective 4D formulation. In this section, we first dimensionally reduce the 6D action by truncating it onto the class of 4D fields that evolve during the cosmologies explored later. We then show that this dimensional reduction is actually a {\em consistent truncation} \cite{DUFF1985355}, which means solutions to the field equations obtained by varying the dimensionally reduced 4D action actually also exactly satisfy {\em all} of the higher-dimensional classical field equations (rather than just satisfying them approximately, as a low-energy approximation). 

\subsubsection{Consistent truncation of the action}

We first derive the dimensionally reduced 4D action obtained by evaluating the 6D action at the truncation 
\be
  \exd \hat s^2 = \hat g_{\mu\nu} (x)\, \exd x^\mu \exd x^\nu + b^2(x) \, \gamma_{mn}(y) \, \exd y^m \exd y^n \,,
 \ee
and $\phi = \phi(x)$. Following \cite{Burgess:2015lda} we also include a dual 4-form field-strength, 
\be
  F_{\mu\nu\lambda\rho} = \frac12 \, \hat\epsilon_{\mu\nu\lambda\rho mn} F^{mn} = f \, \hat\epsilon_{\mu\nu\lambda\rho} \,,
\ee
which does not propagate, but which serves to bring the news to 4D about the quantization condition, eq.~\pref{quantiz}, about which the 4D theory otherwise would have no way of knowing \cite{Polchinski:1995sm}.

Direct substitution of this {\it ansatz} into the 6D action and integrating over the extra dimensions gives
\be
S = -\int \exd^4 x \,4\pi b^2\sqrt{-\hat g} \left\{ \hat g^{\mu\nu} \left[ \f {\hat \cR_{\mu\nu}} {2\kappa^2} - \f 2 {b^2} \, \partial_\mu b \, \partial_\nu b + \f 1 2  \pd_\mu \phi \, \pd_\nu \phi \right] - \f 2 {b^2}   + \f {f^2}2 + V(\phi) \right\} , \label{intAction}
\ee
where the quantities $\sqrt{-\hat g}$ and $\hat \cR_{\mu\nu}$ are the volume density and Ricci tensor for the four-dimensional metric, $\hat g_{\mu\nu}$, and contracted indices run only over the four dimensions with coordinates $x^\mu$. 

The kinetic terms can be put into canonical form through the following field redefinitions. For the extra dimensional radius, we take
\bal \label{radius}
  b &= \bnow \, e^{\psi/2\MP} \,, 
\eal
where 
\be
\MP^2 = \frac{4\pi \bnow^2}{\kappa^2} =: 4 \pi \bnow^2 M_{(6)}^4 \,,
\ee
is the Planck mass and the constant $\bnow = b(\hat t_\star)$ is the present size of the extra dimensions --- so $\psi_\star = \psi(\hat t_\star) = 0$. 4D Einstein frame (4DEF) is achieved through the Weyl rescaling
\bal
g_{\mu\nu} &= e^{\psi/\MP} \hat g_{\mu\nu} \,, \label{change}
\eal
where the condition $\psi(\hat t_\star) = 0$ ensures that transforming to 4D Einstein frame does not also involve a change of units (at the present epoch).

We are led in this way to the 4D action:
\bal
S = - \int \exd^4  x\sqrt{-{g}}\left\{  g^{\mu\nu} \left[ \f {\MP^2}{2}\cR_{\mu\nu} + \f 1 2 \, \partial_\mu \psi \, \partial_\nu \psi  + \f 1 2 \, \partial_\mu \varphi \, \partial_\nu \varphi \right] + W(\varphi,\psi) \right\}, \label{4Daction}
\eal
where we canonically normalize $\varphi := \sqrt{4\pi} \, \bnow\, \phi$. The scalar potential for the two 4D scalar fields, $\varphi$ and $\psi$, is
\bal
W(\psi, \varphi) := 4\pi \bnow^2 e^{-\psi/\MP} U(\varphi) - \f {\MP^2}{\bnow^2}e^{-2\psi/\MP} +  \f {2\pi\mathfrak{f}^2}{\bnow^2}  e^{-3\psi/\MP}, \label{vtot}
\eal
where $U(\varphi) = V(\phi)$ and so $V(\phi) = V_0 \, e^{\kappa \phi}$ implies $U(\varphi) = V_0 \, e^{\kappa\varphi/(4 \pi \bnow)} = V_0 \, e^{\varphi/\MP} $ (say). 

The field equations for this 4D system reduce for homogeneous, spatially flat configurations in a 4DEF FRW metric, 
\be
 \exd  s^2 =  g_{\mu\nu} \exd x^\mu \exd x^\nu =  - \exd  t^2 +  a^2( t ) \,\delta_{ij} \exd x^i \exd x^j \,,
\ee
and $\varphi = \varphi(t)$ and $\psi = \psi(t)$ to:
\bal
\ddot\varphi + 3 H \dot\varphi + \frac{\partial W}{\partial \varphi}  &= 0, \nn\\
\ddot\psi + 3 H \dot\psi + \frac{\partial W}{\partial \psi}  &= 0, \,\, \text{and} \label{4D:EOM}\\
\f {\dot\varphi^2}{2} + \f {\dot\psi^2}{2} + W &= 3\MP^2  H^2 \,, \nn
\eal
where $ H := \dot a /  a$ and over-dots refer to derivatives with respect to ${t}$.

An important aspect of this 4D model is that these field equations exactly reproduce the behaviour of the full 6D system (as opposed to being just a low-energy approximation). That is, as we show in Appendix \ref{App:Consistency}, eqs.~\pref{4D:EOM} are completely equivalent to eqs.~\pref{6EOM3} and \pref{EE}, which themselves suffice to solve all of the 6D field equations (due to the generality of our symmetry {\it ansatz}). In this sense, the truncation described above is a {\em consistent truncation} \cite{DUFF1985355}, and homogeneous solutions of the 4D equations can be dimensionally oxidated to obtain consistent solutions to the full 6D field equations. In consequence, we can trust these solutions even in situations where an effective 4D description normally breaks down, such as when the Hubble scale is greater than the KK mass scale: $ H >  m_\kk$.\footnote{Notice that although $\hat m_\kk \simeq 1/b$ in the original 6D metric, the transformation to 4DEF converts it to $ m_\kk \simeq \bnow/b^2$, which ensures $ m_\kk/ H$ varies with time in the solutions considered later in the same way as does $\hat m_\kk/ \hat H$.}

\subsection{Solutions}
\label{subsec:Analytic}

Before presenting cosmologies obtained by integrating the above field equations numerically we pause first to describe two kinds of simple analytic solutions, since these often capture the main features of the numerical solutions in different regimes.

One simple solution that does {\em not} arise as a solution in the situations we explore is 6D de Sitter space. This solution is never found because it does not share the symmetries of our metric {\it ansatz} \pref{metric}. It would have done so (as we have verified explicitly) if the 2D and 3D subspaces were both curved and if one of these was allowed to be fibred over the other (i.e., to be warped), so that they can combine into a 5-sphere for some values of the parameters. 

\subsubsection{Radius stabilization}
\label{subsection:RadStab}

A virtue of the system under study is that it allows a simple stabilization of the radius of the 2-sphere \cite{Aghababaie:2002be,Salam:1984cj,Braun:2006se}, using what is perhaps the earliest example of flux stabilization \cite{Giddings:2001yu,Dasgupta:1999ss}. We briefly describe this solution here in order to identify the parameter range needed for it to exist, and to establish when this is consistent with our later cosmological solutions.

For the stabilized solution, we take $b=\bnow$ and $\phi = \phi_\star$ to be constants, with $V(\phi)$ minimized at the value $V_\star = V(\phi_\star)$. We take the 4D metric, $\hat g_{\mu\nu}$, to be maximally symmetric with curvature set by Einstein's equations. For de Sitter and flat solutions, we may take $\hat H = {\hat a'}/\hat a$ to be constant (and so ${\hat a''}/\hat a = \hat H^2$), in which case the 6D Einstein equations reduce to:
\be \label{6D:static}
 \frac{2}{\bnow^2} = \kappa^2 \left[ \f {3\mathfrak{f}^2}{2\bnow^4}  + V_\star \right] \qquad \hbox{and} \qquad
  6 \hat H^2 = \kappa^2 \left[- \f {\mathfrak{f}^2}{2\bnow^4}  + V_\star \right] \,. 
\ee

\begin{figure}[htpb]
\centering 
\includegraphics[width=0.9\textwidth]{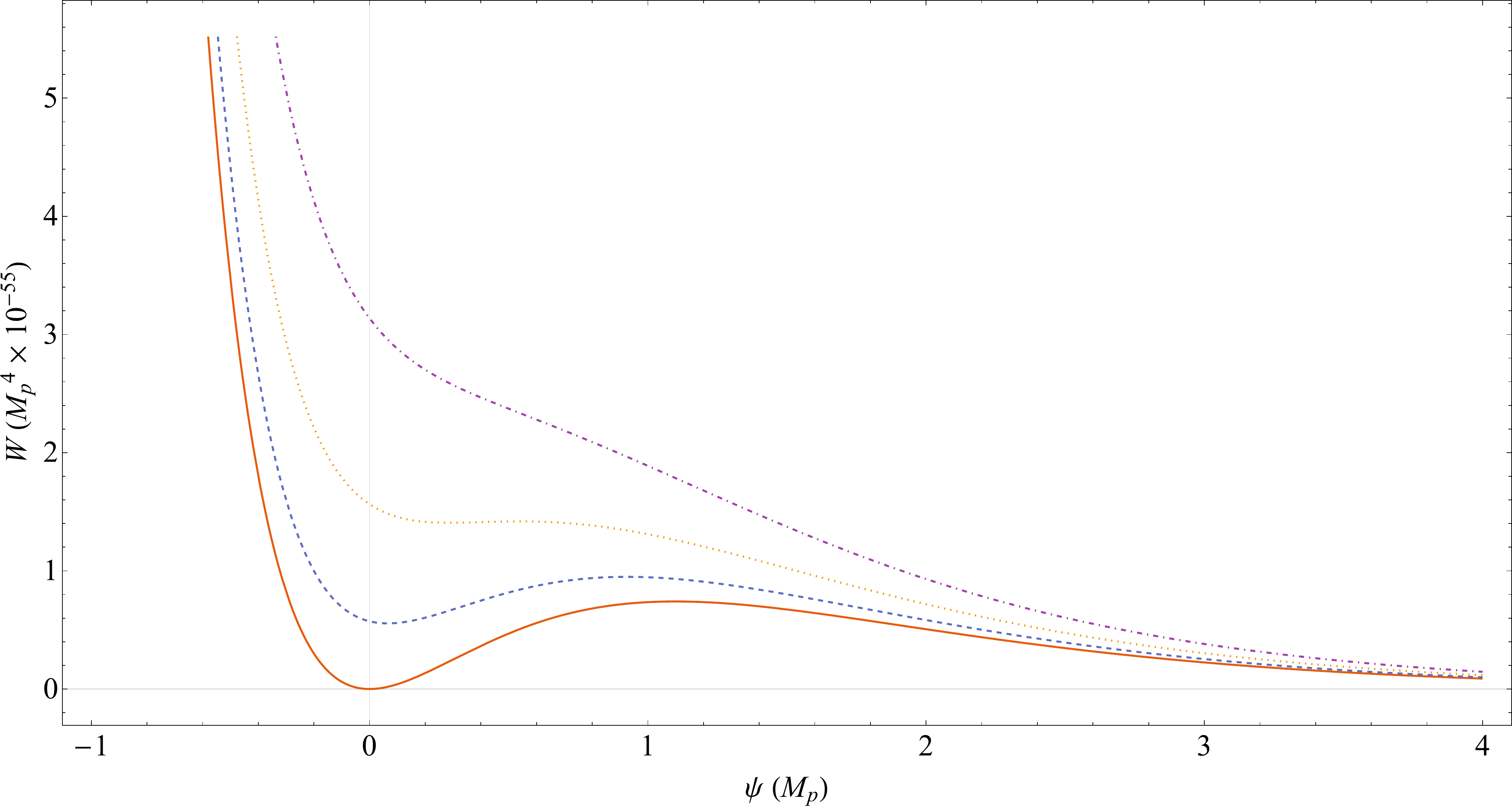}
\caption{Plot of $W(\psi)$ vs $\psi$, both in 4D Planck units, for a few fixed values of $\varphi$. This is the 4D equivalent of the discussion in section \ref{subsection:RadStab}. Notice that a minimum only exists for the radion for certain values of $U(\varphi)$ (hence $\varphi$). It is also important to highlight generically how shallow the well to trap $\psi$ is, when the minimum exists.}
\label{radStab}
\end{figure}

The second of these shows that the solution is 4D de Sitter when $V_\star > \mathfrak{f}^2/2\bnow^4$, but is flat if $V_\star$ is tuned to satisfy $\mathfrak{f}^2 = 2 V_\star \bnow^4$. On the other hand, the first of the above equations can be solved for the size of the extra dimensions in terms of $\mathfrak{f}$ and $V_\star$, with solutions
\be \label{broot}
 \frac{3\kappa^2 \mathfrak{f}^2}{2b_\pm^2} = 1 \pm \sqrt{1 - \frac32 \, (\kappa \mathfrak{f}^2)(\kappa^3 V_\star)} \,.
\ee
As we shall see, $b_+$ is a local minimum of the potential for $b$ while $b_-$ is a local maximum, and the above formula shows that $b_+ < b_-$. The minimum and maximum coalesce into an inflection point when $\kappa^4 \mathfrak{f}^2 V_\star = \frac23$, and no stationary solutions exist within our {\it ansatz} at all for finite $b$ once $\kappa^4 \mathfrak{f}^2 V_\star > \frac23$. An example of how this plays out in the 4D potential $W$ is shown in Figure \ref{radStab} for various values of $V_\star$. 

The conditions for trusting classical solutions become sharper once \pref{broot} is used. As we have seen, semiclassical methods require $\kappa \hat H^2 \ll 1$ and $\kappa/\bnow^2 \ll 1$. This requires small stress energy $\kappa^3 V_\star \ll 1$ and $\kappa^3 F_{mn}F^{mn} \sim \kappa^3 \mathfrak{f}^2/\bnow^4 \sim 1/(\kappa \mathfrak{f}^2) \ll 1$, where the last inequality uses \pref{broot} in the form $\kappa/\bnow^2 \sim 1/(\kappa \, \mathfrak{f}^2)$. Stabilizing the extra dimensions with semiclassical reasoning requires we choose $\kappa \mathfrak{f}^2 = n^2\kappa/(4e^2) \gg 1$, and stabilized extra dimensions are possible in the regime 
\be
 \kappa^3 V_\star \lsim \frac{1}{\kappa \mathfrak{f}^2} \sim \frac{\kappa}{\bnow^2} \ll 1 \,. \label{stab}
\ee 
In particular, the flux stress-energy and 2D curvature are larger than or comparable to the potential when the field $b$ is stabilized.

\subsubsection{Power-law solutions}

Our numerical search for solutions finds many regions where the evolution takes a simple approximate scaling form. It is useful displaying these approximate scaling solutions explicitly so as to more simply explore their properties, as we do in this section.

Scaling solutions are easiest to interpret from the 4D perspective, and typically arise when the potential $W(\psi, \varphi)$ is dominated by a single exponential term \cite{Copeland:1997et}. Two classes of solutions are of most interest in what follows, depending on the relative importance of different terms in the total potential that dominate. In order to quantify these we write $W = W^{(\varphi)} + W^{(\text{c})} + W^{(f)}$, with 
\be
 W^{(\varphi)} = 4\pi \bnow^2 \, U(\varphi) \, e^{-\psi/\MP} \,, \qquad 
 W^{(c)} = - \left( \f {\MP^2}{\bnow^2} \right) e^{-2\psi/\MP}
\ee
and
\be
 W^{(f)} = \left( \f {2\pi \mathfrak{f}^2}{\bnow^2} \right) e^{-3\psi/\MP} \,.
\ee
$W^{(\varphi)}$ arises from the inflaton potential, $W^{(c)}$ is the contribution due to the curvature of the extra dimensions, and $W^{(f)}$ is the contribution due to the extra-dimensional flux. 

When seeking scaling solutions, we assume only one of the exponentials of the 6D potential \pref{6Dpot} dominates, so we can approximate 
\be \label{scale:pot}
   V(\phi) \simeq V_0 \, e^{\lambda \phi/\MP} \,,
\ee
in the regime of interest. We then seek power-law scaling solutions of the form
\bal
\frac{\varphi}{\MP} &= \frac{\varphi_0}{\MP} + p_1 \ln \left({t}/{t}_0\right), 
\nn\\
\frac{\psi}{\MP} &= \frac{\psi_0}{\MP} + p_2 \ln \left({t}/{t}_0\right), \label{scale:psi}\\
 a &=  a_0 \left( \f { t} { t_0} \right)^{ \alpha}, \nn
\eal
from which we also know 
\be
 H = \f { \alpha} { t} =  H_0 \left( \frac{b_0}{b} \right)^{2/p_2} \,.
\ee 

These scaling solutions provide accelerated expansion when $ \alpha > 1$ and (as is always true for power-law solutions) the slow-roll parameters are given by
\be
   \epsilon := - \frac{ \dot H}{ H^2} = \frac{1}{ \alpha} \qquad \hbox{and so} \qquad
   \eta := \frac{ \dot\epsilon}{ H  \epsilon} = 0 \,.
\ee
The scaling behaviour alone also suffices to determine the amount of time spent doing so, with the number of $e$-foldings given by
\be \label{epsiloneq}
 N_e := \ln \left( \frac{ a_f}{ a_0} \right) = \int_{ t_0}^{ t_f} \exd  t\,\,  H = \alpha\ln \left( \frac{{t}}{{t}_0} \right) = \f { \alpha} {p_2}\left( \frac{\psi_f - \psi_0}{\MP}  \right) = \frac{2  \alpha}{p_2} \ln \left( \frac{b_f}{b_0} \right) \,,
\ee
where the second-last equality uses \eqref{scale:psi} and the last equality trades the canonical field $\psi$ for the geometrical quantity $b$. Amusingly, if $b_f/b_0 \sim 10^{15}$  during such a scaling regime,\footnote{Such as would happen if $b$ ran from the Planck scale to the electroweak scale, or from the electroweak scale to the micron scale.} then the above relation predicts the `usual' three spatial dimensions expand by 
\bal 
 N_e \sim 70 \left( \f {  \alpha} {p_2} \right) \,, \label{pl:efolds}
\eal
$e$-foldings. This potentially provides a novel connection between the number of inflationary $e$-foldings and the current size of two extra dimensions, with 70 $e$-foldings emerging from natural (order-unity) choices for the powers $ \alpha$ and $p_2$.

Another combination of later interest, whose time-dependence is fixed in these solutions, is the ratio of the Hubble scale to the KK-mass scale, which, in the 4DEF, are given by $ H$ and $ m_\kk \simeq \bnow/b^2$. Their ratio therefore depends on time as
\be \label{HvsKK}
 \frac{ H}{ m_\kk} = \frac{ H_0}{ m_{\kk_0}}  \left( \frac{b_0}{b} \right)^{(2/p_2)-2} \,.
\ee

The values for powers like $\alpha$ follow from the equations of motion and their precise values depend on which of the terms in $W$ dominate. We identify the following particularly interesting cases:

\subsubsection*{Attractor Solution} 
\label{attractor}

This solution is obtained when $\abs{W^{(\varphi)}} \gg \abs{W^{(c)}}, \abs{W^{(f)}}$. In this case, the equations of motion \eqref{4D:EOM} reduce to
\bal
\f {p_1(3\alpha - 1)} { t^2} &= \frac{\lambda  U_0}{\MP^2} \left( \f { t} { t_0} \right)^{-(\lambda p_1 + p_2)}, \nn\\
\f {p_2(3\alpha - 1)} { t^2} &= \f { U_0} {\MP^2}  \left( \f { t} { t_0} \right)^{-(\lambda p_1 + p_2)}, \\
\f {6\alpha^2 - p_1^2 - p_2^2}{ t^2} &= \frac{2 U_0}{\MP^2} \left( \f { t} { t_0} \right)^{-(\lambda p_1 + p_2)},\nn
\eal
where $ U_0 := 4\pi \bnow^2 \, V_0 \exp[-(\lambda \varphi_0 + \psi_0)/\MP]$. Equating exponents of time and the overall coefficients of those powers, we find the powers:
\be \label{att:alpha}
 \alpha = \f {2} {1 +  \lambda^2} 
\,, \qquad 
p_1 =  \alpha\lambda 
\,, \qquad
p_2 = \alpha  \,, 
\ee
and a relation amongst the coefficients
\be \label{att:coef}
 t_0^2 = \f {\left( 5 - \lambda^2\right) \alpha ^2\MP^2} {2 U_0} \,.
\ee

Since $ \alpha \le 2$ for all real $ \lambda$, \pref{epsiloneq} implies a lower bound $ \epsilon \ge \frac12$, so the slow roll is never quite that slow. Furthermore, $ \alpha = p_2$ and so the number of $e$-foldings is $ N_e = 2 \ln (b_f/b_0)$. Finally, because $p_2 =  \alpha$ we have $2/p_2 = 2 \epsilon \ge 1$ and so $ H$ is a strictly falling function of time. It need not fall faster than $ m_\kk$, however, and in particular for $ \epsilon = \frac12$, \pref{HvsKK} shows that $ H/ m_\kk \propto (b_0/b)^{2( \epsilon - 1)}$, which {\em grows} with $b$ for the accelerating solutions (for which $\frac12 \le  \epsilon < 1$).

What makes this scaling solution particularly special is that it is an attractor: when initial conditions are perturbed, nearby trajectories tend to converge towards this solution again at late times. This is shown in detail in Appendix \ref{App:Stability}, where the linearized equations are solved and show that perturbations converge like a calculable power of time. This attractor behaviour is evident in our numerical solutions, which are often drawn towards this scaling solution starting from a broad variety of initial conditions. Once established, the attractor solution tends to persevere until one of the assumptions underlying its existence begins to fail. Chief amongst these is the assumption that $W$ is dominated by a particular exponential term. For instance, eventually motion of the scalar field, $\phi$, finds its minimum at which point $V(\phi)$ of \pref{6Dpot} is no longer dominated by a single exponential.

From the point of view of inflation, what is attractive about this solution is its attractor nature. This makes it relatively insensitive to initial conditions in a way that is not true for many inflationary models. What is unattractive is the lower bound $ \epsilon \ge \frac12$, since this precludes using it to describe primordial fluctuations using the standard mechanism of inflaton vacuum fluctuations. 

\subsubsection*{Slow-roll solution}

There is a second power-law solution exposed in our numerical solutions, which differs from the attractor just described in several ways. The main two of these are: it can allow much smaller values of $ \epsilon$; and it is not an attractor, since the stability analysis of Appendix \ref{App:Stability} reveals a single growing mode for perturbed trajectories. Indeed numerical solutions that start in the present scaling solution generically eventually cross over to the attractor solution just described.

For this solution, we assume the hierarchy amongst terms in $W$ to be $\abs{W^{(\varphi)}} \sim \abs{W^{(c)}} \gg \abs{W^{(f)}}$. In this case, the equations of motion are:
\bea
&&\qquad\qquad\qquad\qquad\qquad \f {p_1(3\alpha - 1)} { t^2} = \frac{ \lambda  U_0 }{\MP^2} \left( \f { t} { t_0} \right)^{-(\lambda p_1 + p_2)} \nn\\
&& \qquad \f {p_2(3\alpha - 1)} { t^2} = \f { U_0}{\MP^2} \left( \f { t} { t_0} \right)^{-(\lambda p_1 + p_2)} -  \f {2} {\bnow^2}\, e^{-2\psi_0/\MP} \left(\f { t} { t_0}\right)^{-2p_2} \\
&& \f {6\alpha^2 - p_1^2 - p_2^2}{ t^2} =\frac{ 2 U_0}{\MP^2} \left( \f { t} { t_0} \right)^{-(\lambda p_1 + p_2)} -  \f {2} {\bnow^2} \, e^{-2\psi_0/\MP} \left(\f { t} { t_0}\right)^{-2p_2}. \nn
\eea
Again, equating the exponents and coefficients of time leads to the following prediction for the powers
\be
p_1 = \f 1 { \lambda} \,, \qquad
p_2 = 1 \qquad \hbox{and} \qquad
 \alpha = \f {1 + \lambda^2} {2\lambda^2} \,, \label{un:alpha}
\ee
and for the coefficients
\be \label{un:coef}
 U_0 = \f {2\MP^2} {\left( 1 - \lambda^2 \right) \bnow^2}  \, e^{-2\psi_0/\MP}\qquad \hbox{and} \qquad
 t_0^2 = \f {\left( \lambda^2 + 3 \right)\MP^2} {2\lambda^4  U_0} \,.
\ee
Expression \pref{un:alpha} for $\alpha$ implies the first slow-roll parameter for this solution is
\be
 \epsilon = \frac{1}{\alpha} =   \f {2\lambda^2} {1 + \lambda^2} 
\ee
and $\eta = 0$, showing that $\epsilon$ can be made arbitrarily small  (as advertised above) by choosing $\lambda$ sufficiently small.

Furthermore, in this case $ \alpha / p_2 =  \alpha = 1/ \epsilon$ and so the number of $e$-foldings is $ N_e = (2/\epsilon) \ln (b_f/b_0)$, which can be large even if $b_f/b_0$ is not. Finally, because $p_2 = 1$, we have $2/p_2 = 2$ and so $ H$ is again a strictly falling function of time. Indeed, it falls at the same rate as $ m_\kk$ and so $ H/ m_\kk$ is time-independent for this solution. 

\section{Inflationary examples}
\label{section:numerics}

We have numerically found three classes of solutions. The first is an example where the radion is trapped almost immediately, and the rest of space exhibits the usual slow-roll inflation. In the second and third classes, we find explicit examples of both scaling solutions described above. Each of these solutions has its own benefits and drawbacks, which we detail below.

\subsection{Cradle Inflation}

\begin{figure}[h]
\begin{center}
\includegraphics[height=70mm]{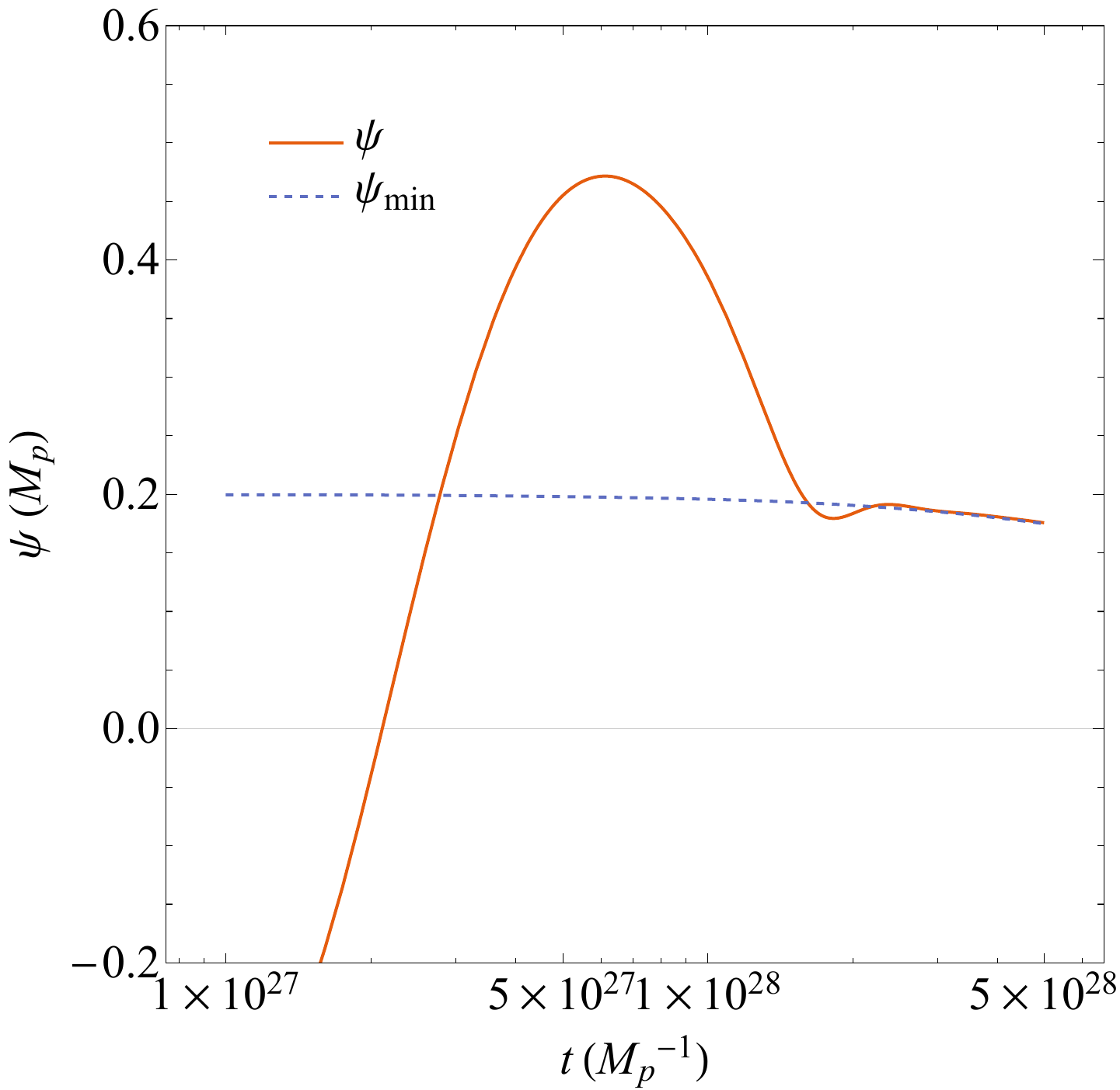}
\qquad
\includegraphics[height=70mm]{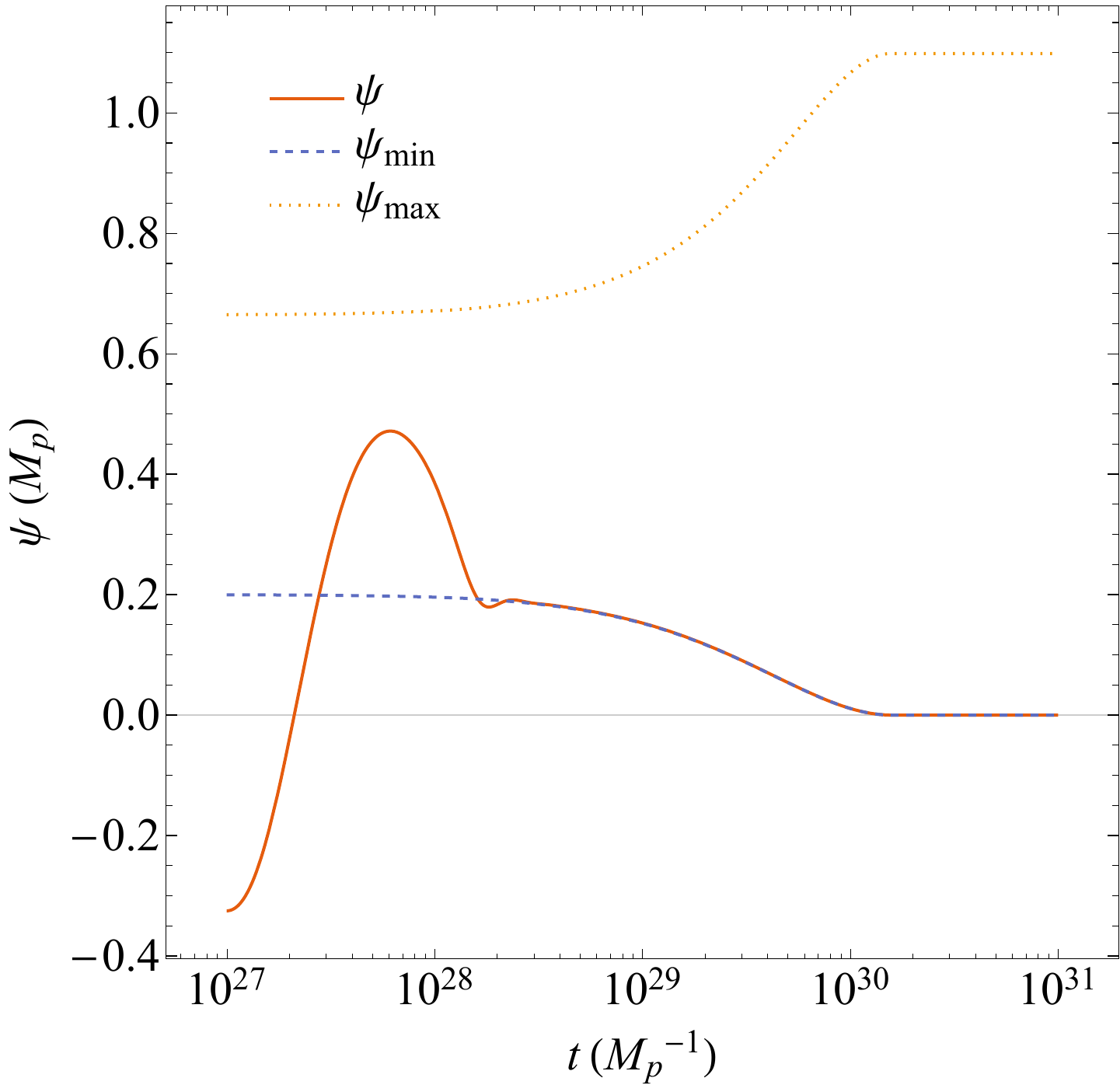}
\caption{Behaviour of the radion in the Cradle scenario. $\psi_{\text{min}}$ and $\psi_{\text{max}}$ correspond to the $+$ and $-$ roots of \eqref{broot}, respectively. Behaviour at early times is enlarged for clarity (left).}
\label{cradle1}
\end{center}
\end{figure}

This solution is found by letting the radion, $b$ or $\psi$, start from close to rest very near to the minimum of its potential.\footnote{We note that from \eqref{broot}, this solution is only available when $V(t_0)$ satisfies $\frac32 \, (\kappa \mathfrak{f}^2)(\kappa^3 V(t_0)) < 1$, since there otherwise doesn't exist a minimum for $b$.} With this choice, the radion simply performs damped oscillations around its local minimum, as shown in Figure \ref{cradle1}, with a damping set by Hubble friction.

But this does not mean that inflation is eternal (because the radion stays at its minimum); nor does it mean that the fields $\varphi$ and $\psi$ remain time-independent after the radion finds its minimum. On the contrary, the potential continues to evolve and eventually ends because $\psi$ finds its local minimum of $W$ while $\varphi$ is still rolling towards its own stable endpoint. Inflation occurs because $\varphi$ is not yet at its minimum and so $W$ is not zero despite $\psi$ being minimized. (After all, it is only when {\em both} fields are minimized that $\Lambda$ tunes $W$ to vanish.) Consequently, $\psi$ is cradled within a local basin and has its energy slowly lowered to zero as $\varphi$ makes its way towards its own final resting place. 

\begin{figure}[h]
\centering
\includegraphics[height=70mm]{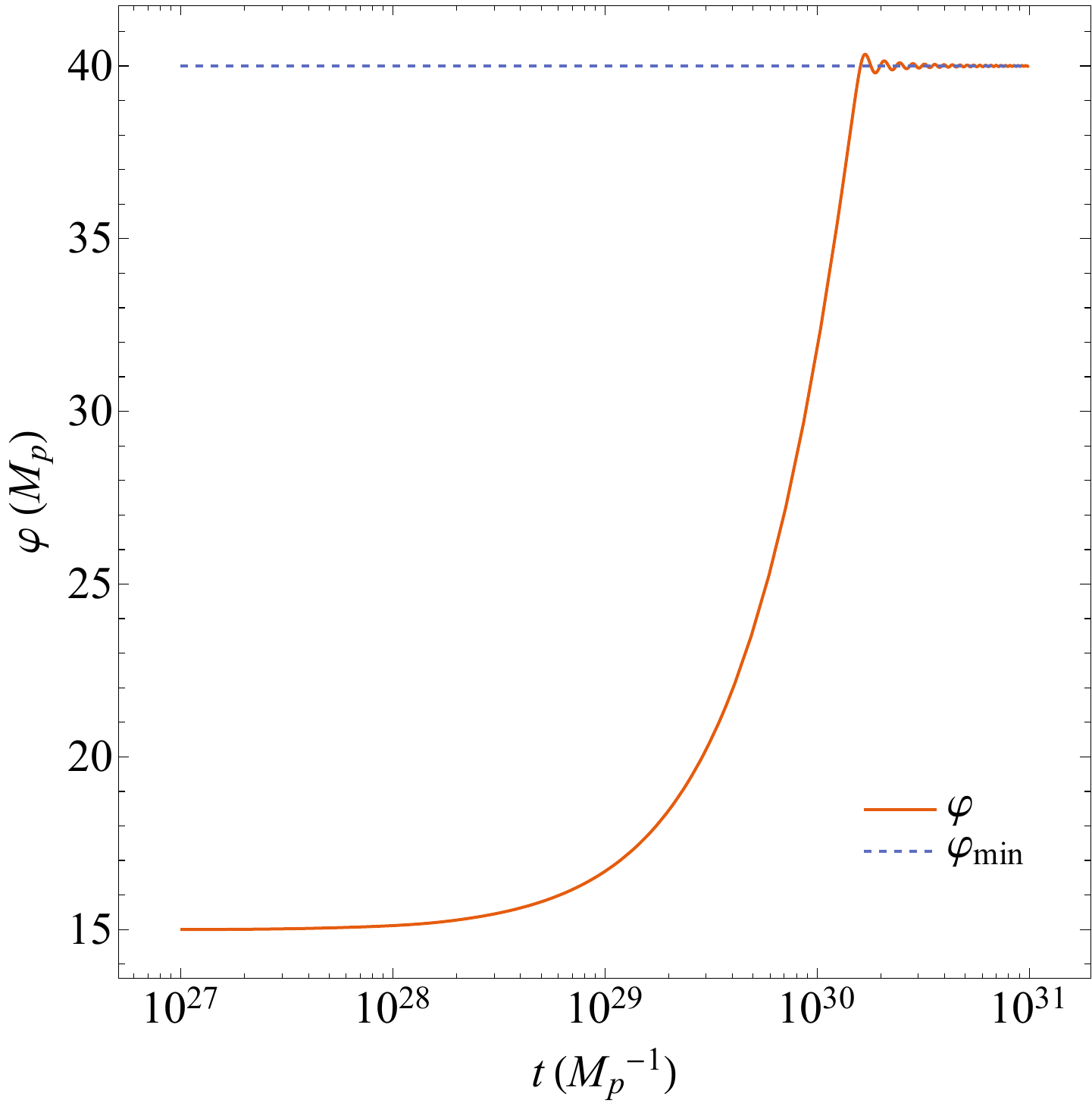}
\qquad
\includegraphics[height=70mm]{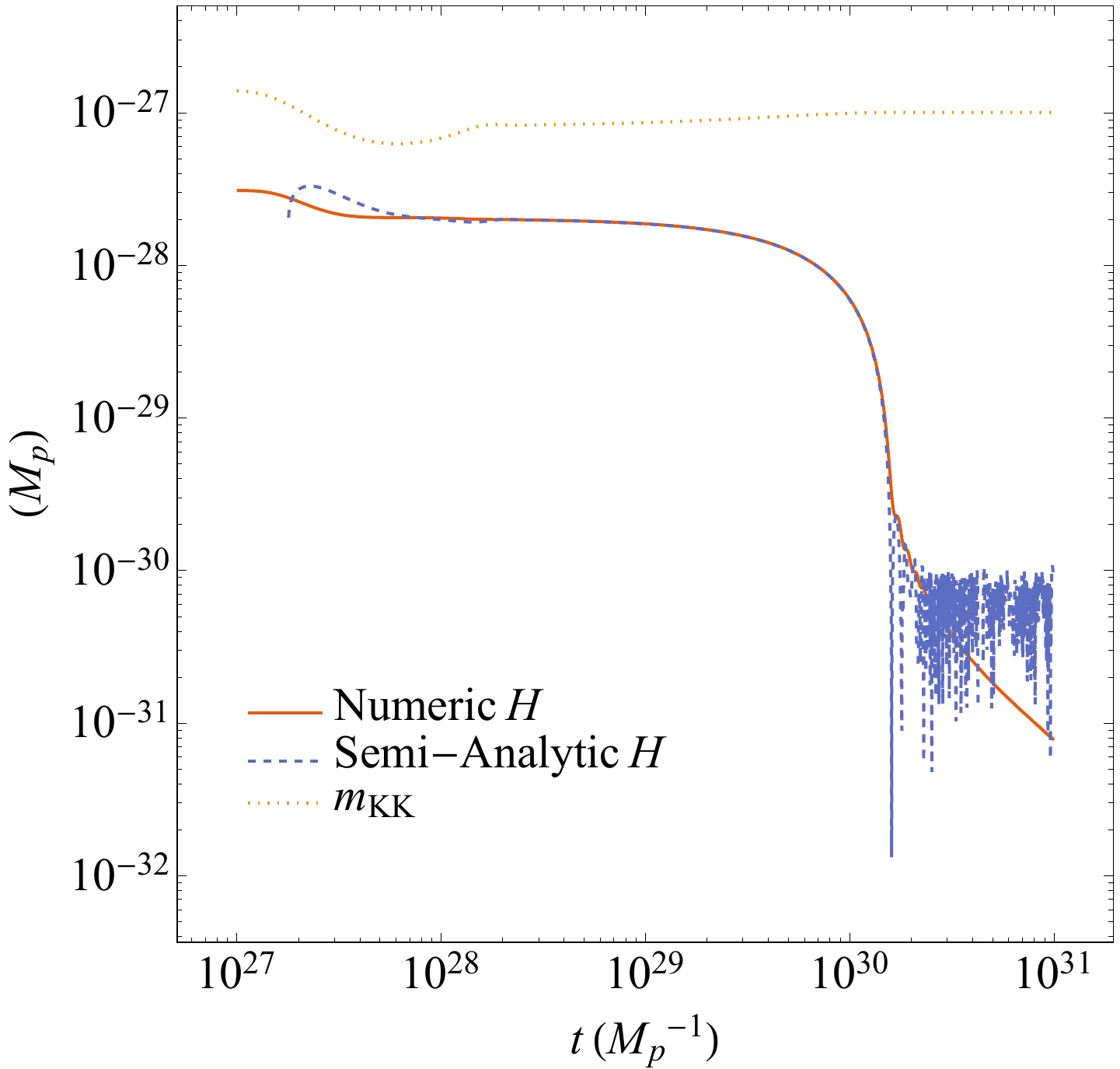}
\caption{4D Inflation in the Cradle scenario. In the second plot, the semi-analytic curve is a plot of equation 2.23 (using A.1 to convert to 4D quantities). The slow-roll inflationary regime occurs where the numeric and semi-analytic curves agree.}
\label{cradle2}
\end{figure}

The evolution of $\varphi$ towards its minimum with $\psi$ pre-trapped is shown in the left panel of Figure \ref{cradle2}. The 4D Hubble parameter during this roll is shown in the right panel of the same figure. Shown also in this panel for comparison is the size of the KK-mass scale, showing that the Hubble rate is smaller than the KK-mass and so the model lies within the regime well-described by a traditional effective 4D description. As a result, we expect that the generation of primordial density fluctuations is likely to go through much as would be expected for any 4D two-field model of this type.

\begin{figure}[h]
\centering
\includegraphics[height=65mm]{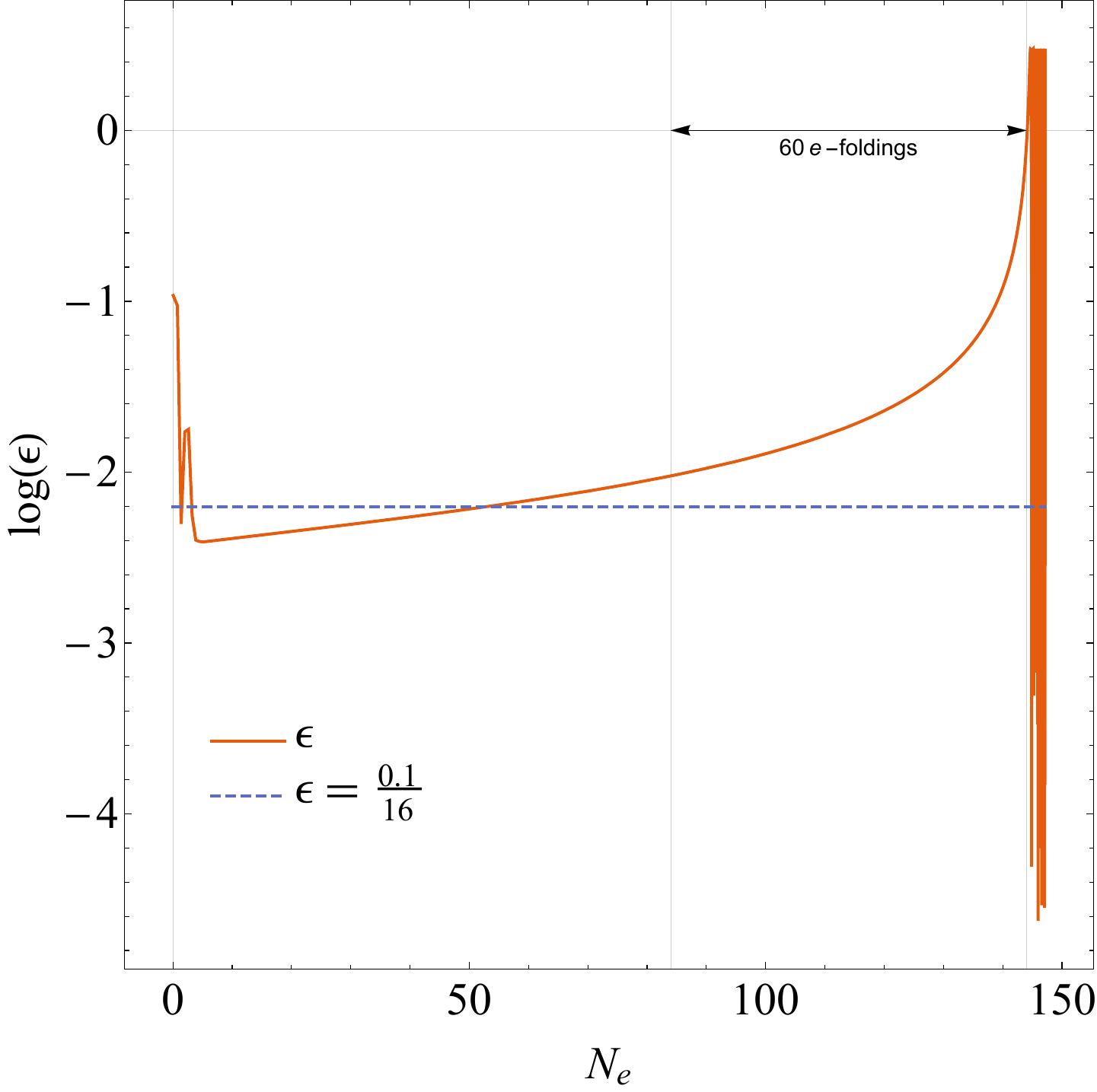}
\qquad
\includegraphics[height=65mm]{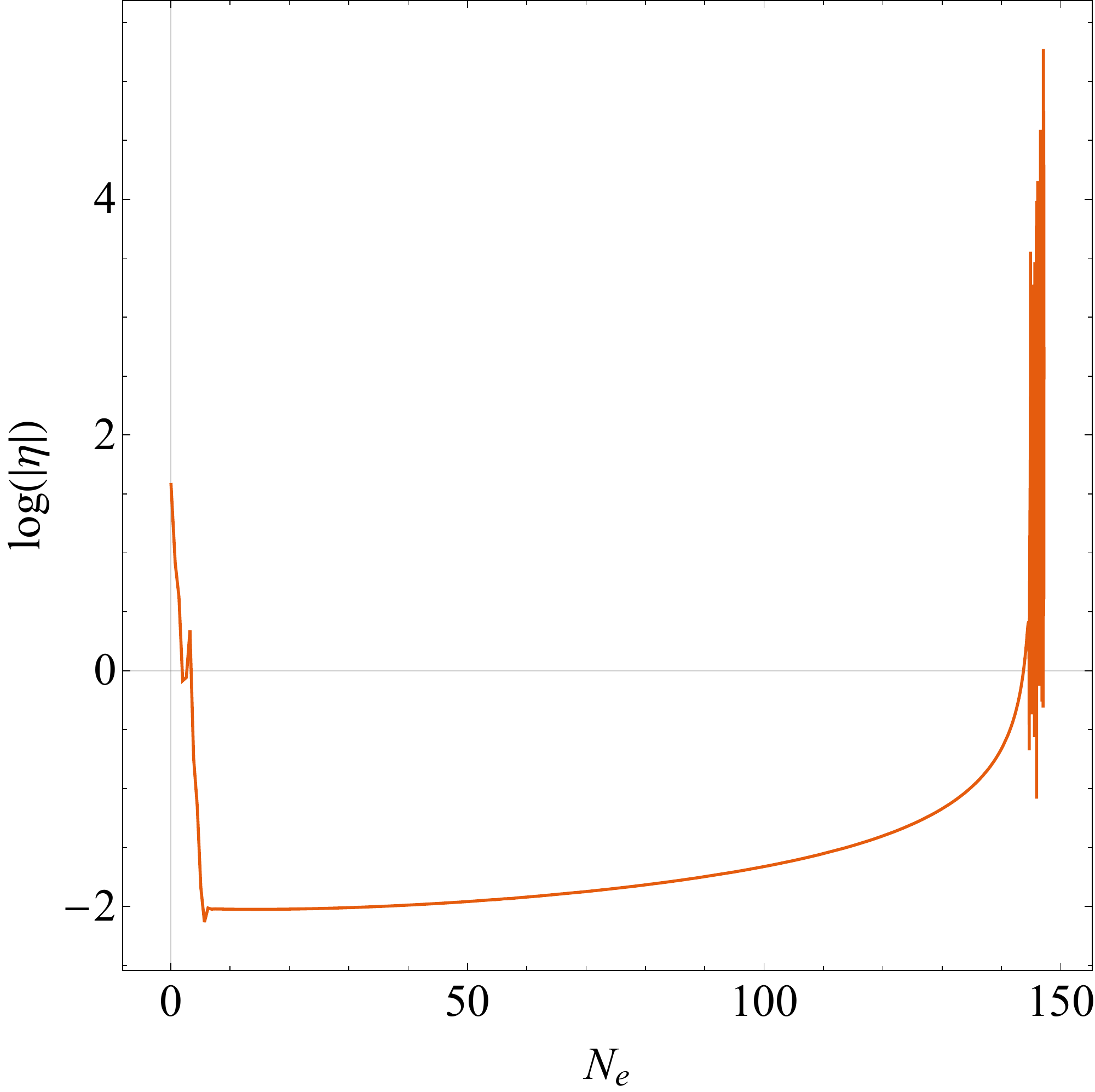}
\caption{Slow-roll parameters $\epsilon$ and $\eta$ during the inflationary regime for the Cradle scenario. They are plotted as a function of $e$-foldings, $N_e$.}
\label{cradle3}
\end{figure}

Since the evolution is driven by the rolling of $\varphi$, the size of the slow-roll parameters can be controlled from the shape of $V(\phi)$ using fairly standard methods, and since $H \ll m_\kk$, we may use the standard 4D prediction 
\be
  n_s = 1 -6 \epsilon + 2\eta \qquad \hbox{and} \qquad r = 16  \epsilon \,. 
\ee
In the representative solution displayed in the figures, the slow-roll parameter $\epsilon$ and $ \eta$ approximately 60 $e$-foldings before inflation's end are $\epsilon \simeq 0.009$, and $\eta \simeq 0.016$, leading to the predictions $n_s \simeq 0.975$ and $r \simeq 0.15$. This value of $r$ is on the high side of the phenomenologically acceptable region \cite{Adam:2015rua} at horizon exit, and (as seen in Figure \ref{cradle3}) lies in the range $0.13 \lsim r \lsim 0.18$ if we take $70 \gsim N_e \gsim 50$ $e$-foldings.  One can imagine seeking better agreement if the post-inflationary physics allowed inflation to last for a longer time, or (better) if the potential $V(\phi)$ is not chosen as the sum of exponentials as in \pref{6Dpot} (which was chosen here more to study the scaling solutions described below).

\subsection{Attractor}

Our second numerical solution demonstrates regimes that behave as does the attractor scaling solution detailed in section \ref{attractor}. This is done by choosing initial conditions that ensure the hierarchy $\abs{W^{(\varphi)}} \gg \abs{W^{(c)}}, \abs{W^{(f)}}$ of the terms in $W$.

\begin{figure}[h]
\centering
\includegraphics[height=72mm]{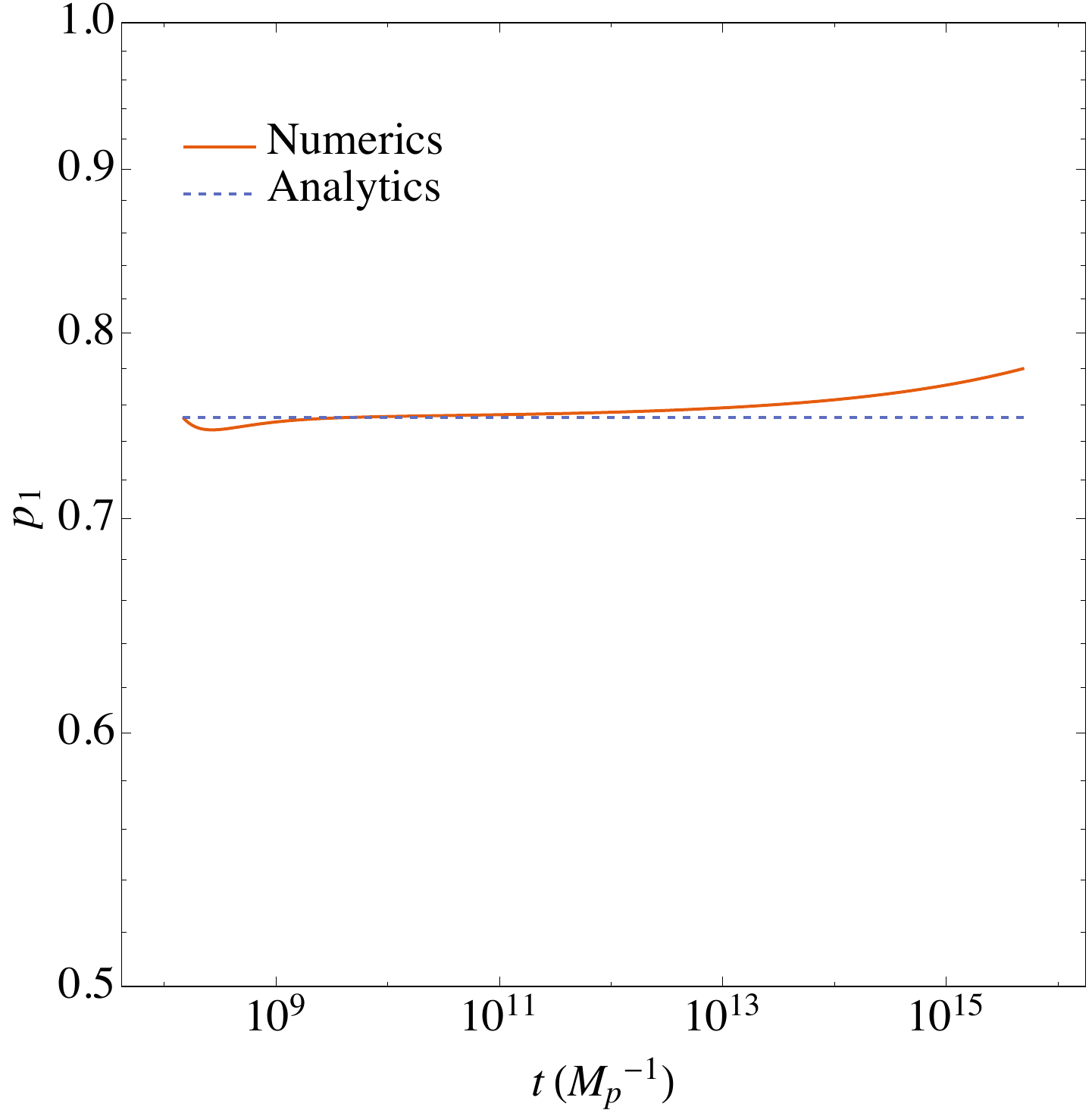}
\qquad
\includegraphics[height=72mm]{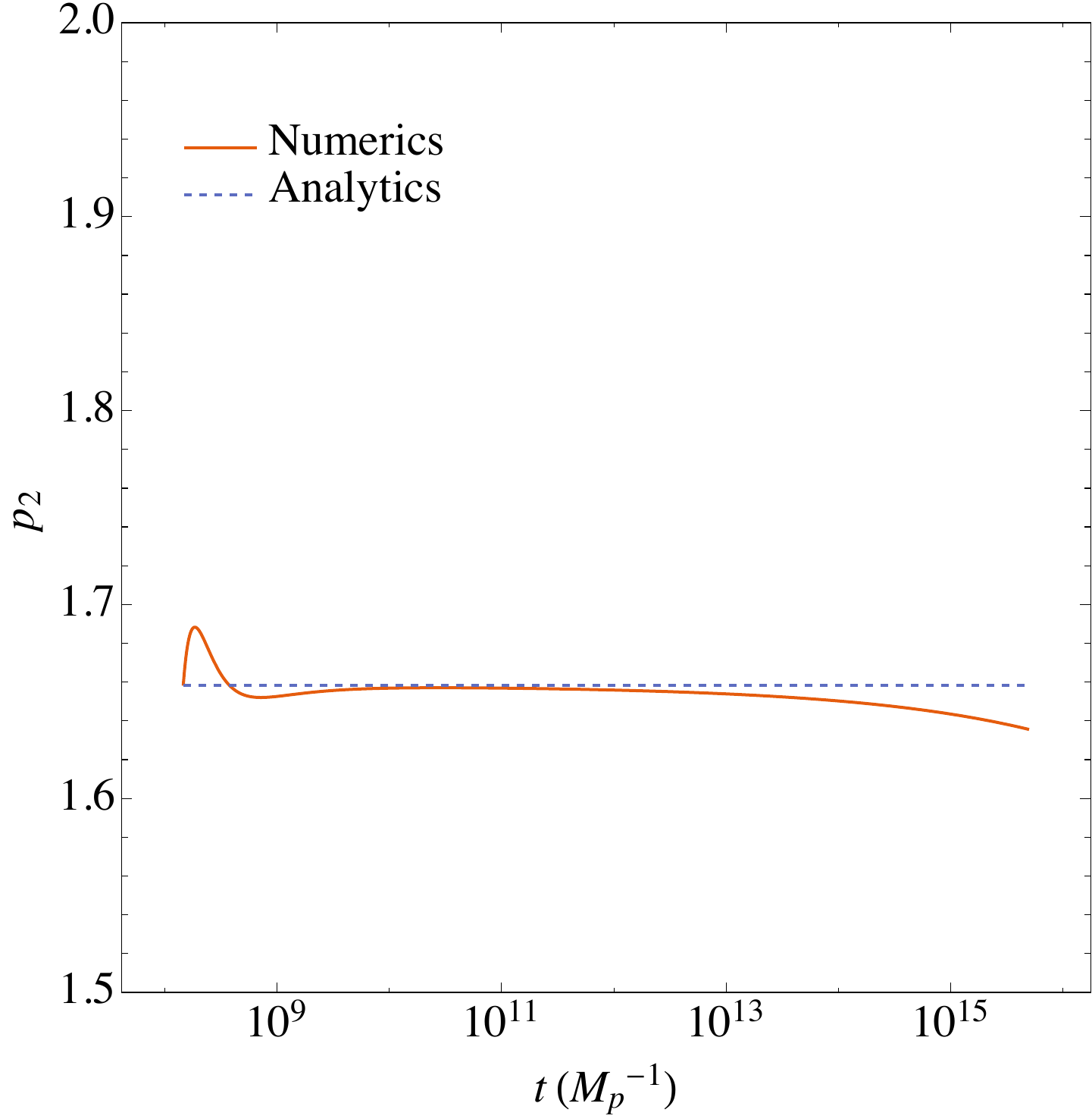}
\caption{Plots of $ t \dot \varphi$ (left) and $ t \dot\psi$ (right) vs $ t$. Reasonable agreement is seen between the numerics and analytics, indicating we are reasonably seeing the attractor scaling solution. Note that the agreement begins to break down as the approximation that $U(\varphi)$ is a single exponential begins to fail.}
\label{att1}
\end{figure}

\subsubsection{Power-Law Behaviour}

\begin{figure}[h]
\centering
\includegraphics[width=0.7\textwidth]{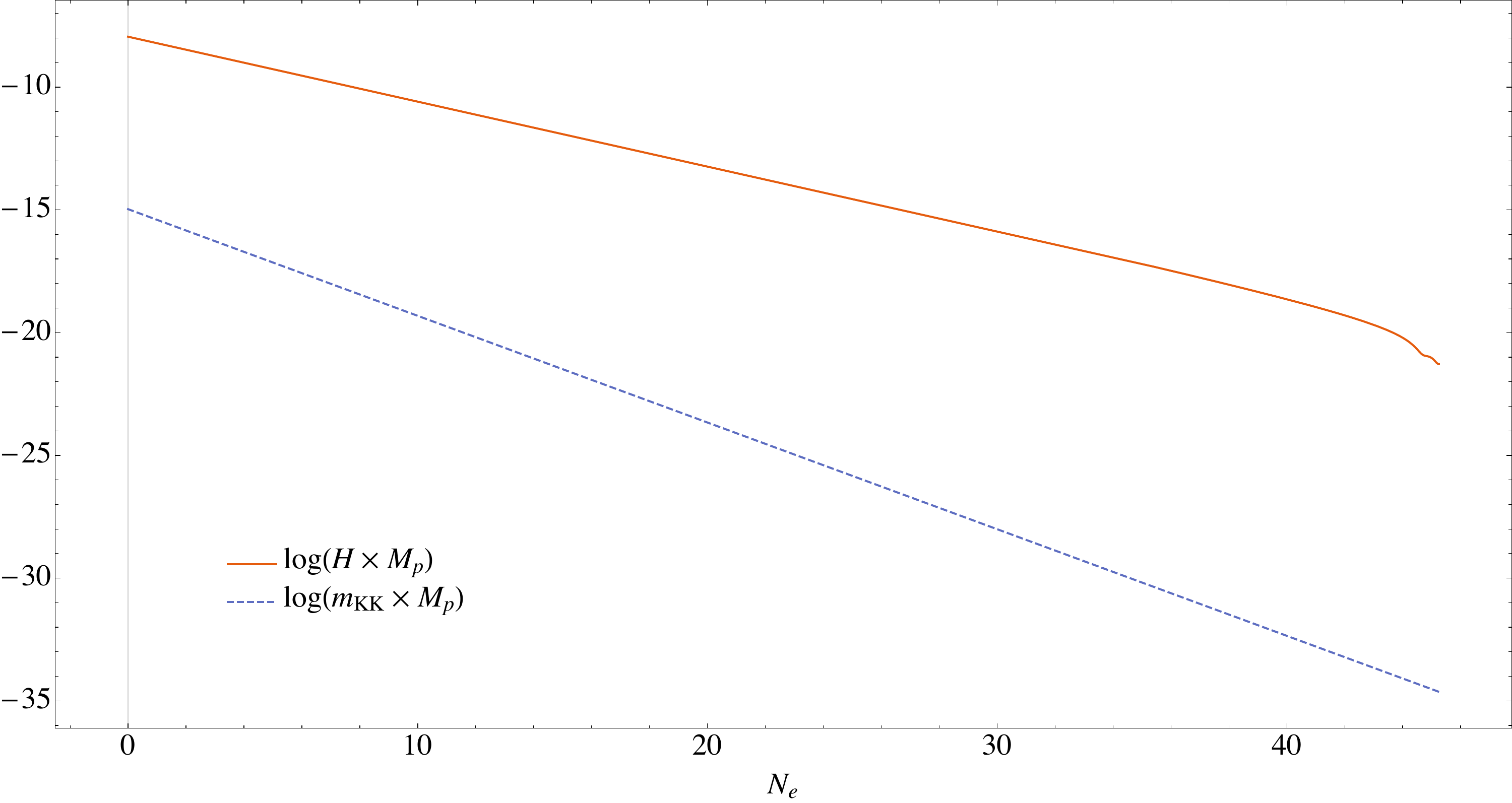}
\caption{Plot of log $ H$ and log $ m_{\kk}$ vs $N_e \propto \hbox{log}\, t$. Notice that we see the ratio $ H/ m_{\kk}$ \emph{increase} with time.}
\label{att2}
\end{figure}

\begin{figure}[h]
\centering
\includegraphics[height=72mm]{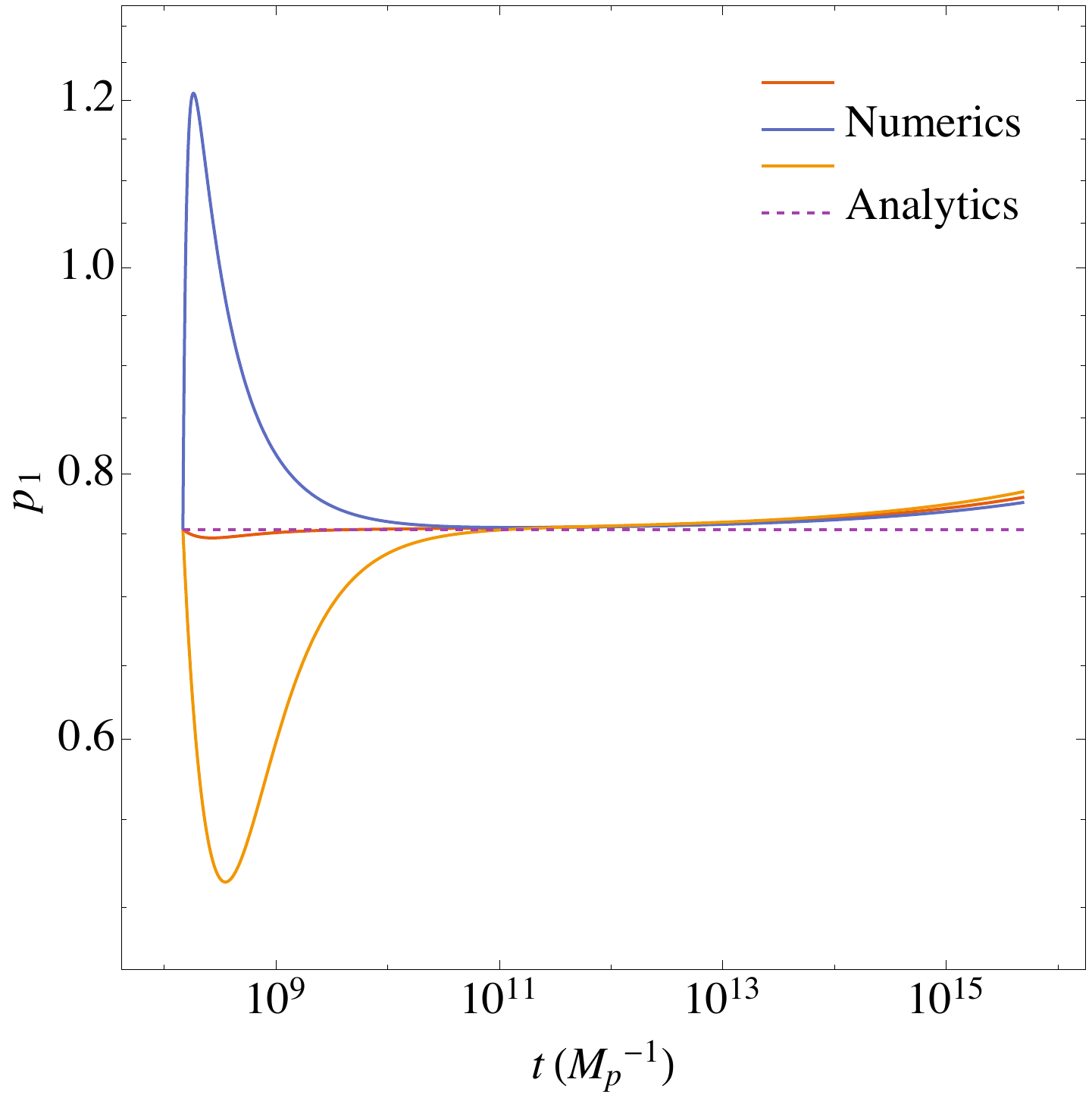}
\qquad
\includegraphics[height=72mm]{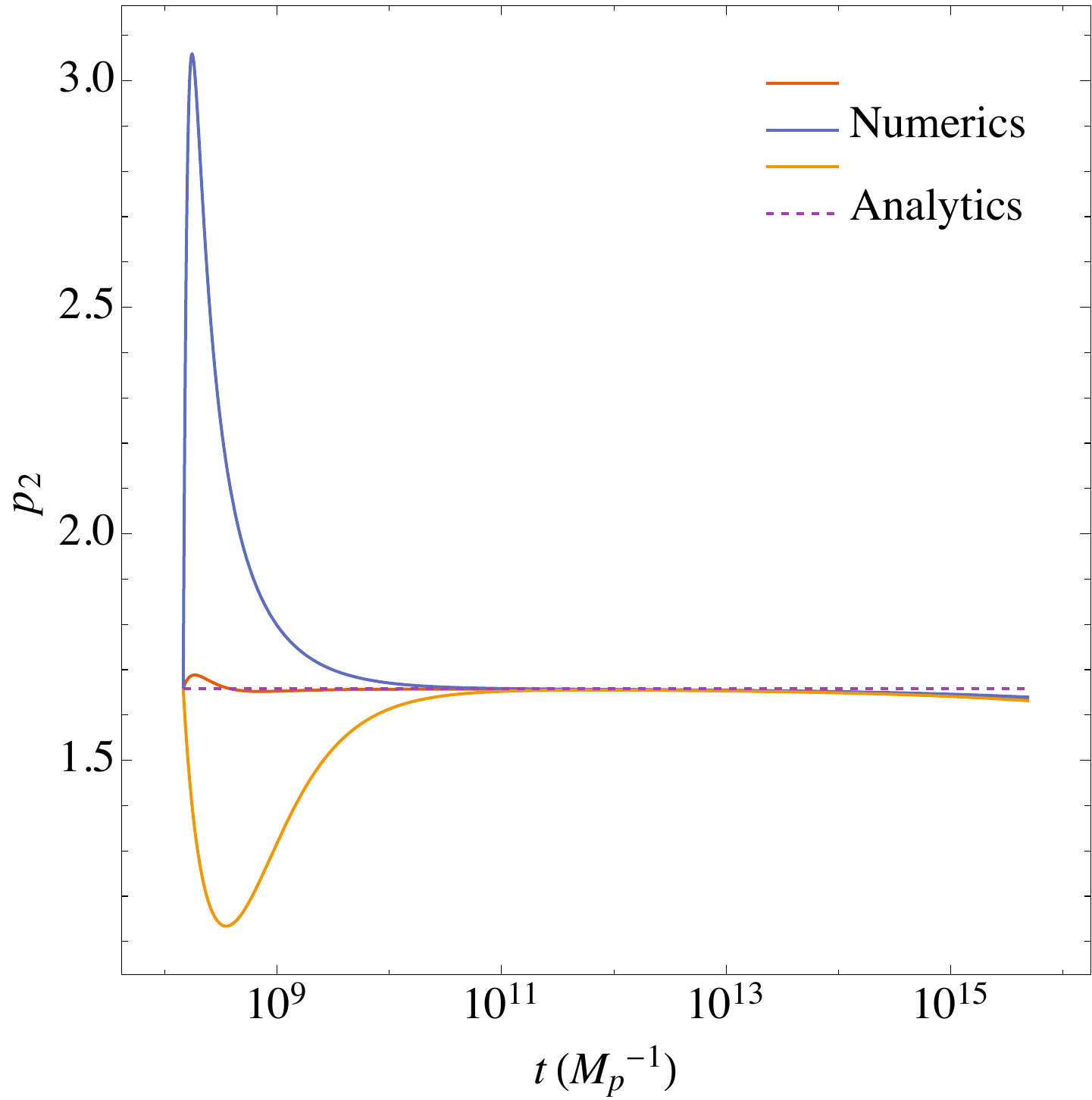}
\caption{Plots of $ t \dot \varphi$ (left) and $ t \dot \psi$ (right) vs $ t$ for several similar initial conditions (solid lines are numerics, dashed lines are analytics). Note that despite an assortment of initial conditions, each solution approaches roughly the same behaviour.}
\label{att3}
\end{figure}

There are several features of the analytical scaling solution that can be used to identify when it provides a good description of a numerical solution. Perhaps the easiest of these is the prediction that the velocities of $\varphi$ and $\psi$ vary as $ \dot\varphi = p_1/t$ and $\dot\psi = p_2/t$. Figure \ref{att1}, shows an example with a regime of good agreement between these predictions and the numerics. Another diagnostic is the time-dependence of $H/m_\kk$ which Figure \ref{att2} shows is also well-captured by numerical integration. The final diagnostic is the attractor behaviour itself: we expect numerical integration starting from slightly different initial conditions should tend to converge towards the attractor solution. In Figure \ref{att3}, we plot the behaviour of the radion and the inflaton for a variety of initial conditions close to the expected scaling solution and see they indeed all approach a common trajectory.

\subsubsection{Trapping}

To profit from modulus stabilization we also seek solutions that trap the radion in its minimum after leaving the scaling regime. Unfortunately, our numerics show that this is actually generically difficult in practice, for two reasons. First, recall from Figure \ref{radStab} that the radion doesn't have a global minimum at finite $\psi$ for all values of $\varphi$. Consequently parameters and initial conditions must be chosen to ensure that exit from the scaling solution occurs where a minimum for $\psi$ exists. Happily, it happens that the defining condition of the attractor scaling solution --- $\abs{W^{(\varphi)}} \gg \abs{W^{(c)}}, \abs{W^{(f)}}$ --- is consistent with what is required for there to exist a nontrivial minimum of $W$ for $\psi$. 

\begin{figure}[h]
\centering
\includegraphics[height=70mm]{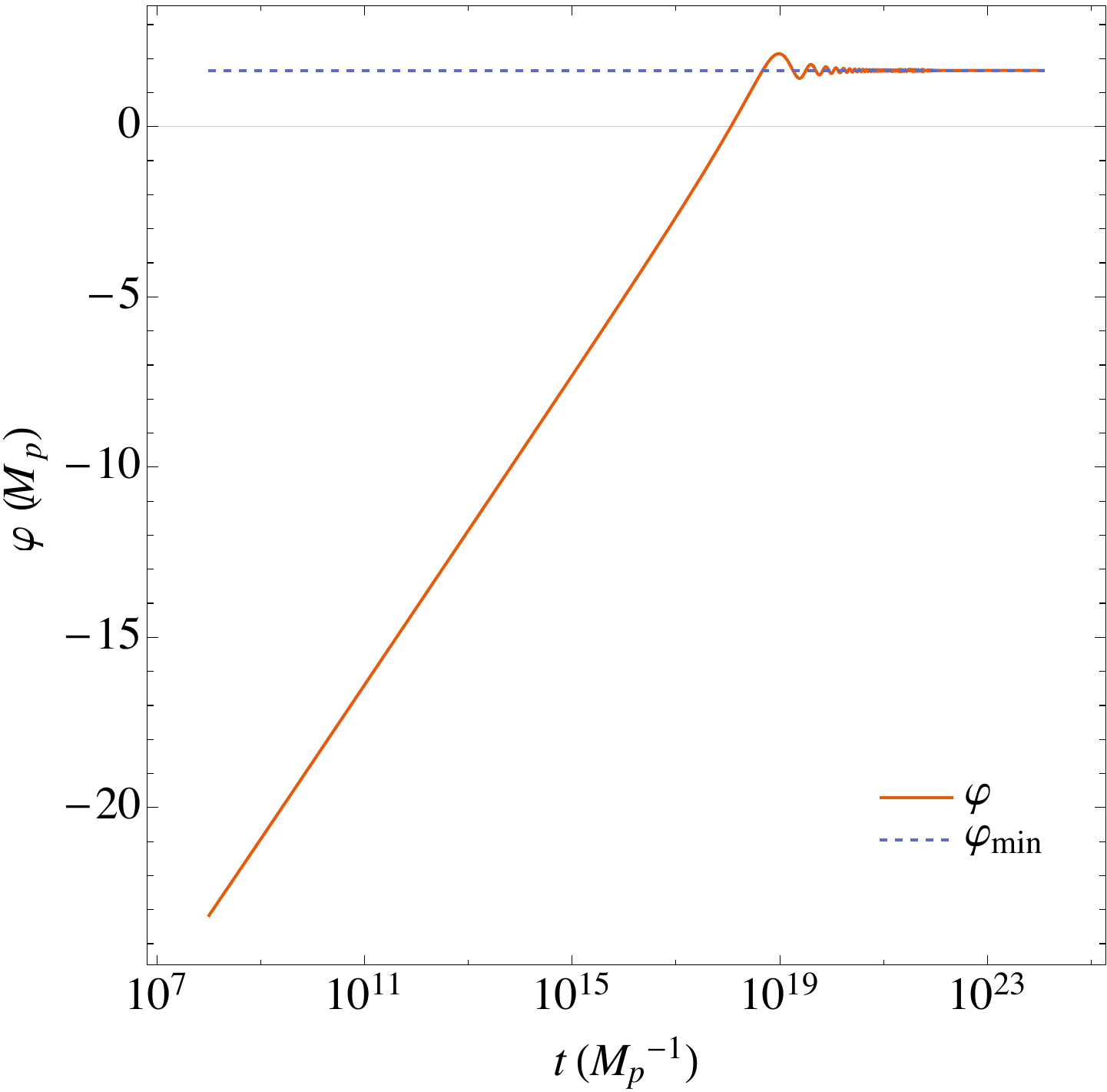}
\qquad
\includegraphics[height=70mm]{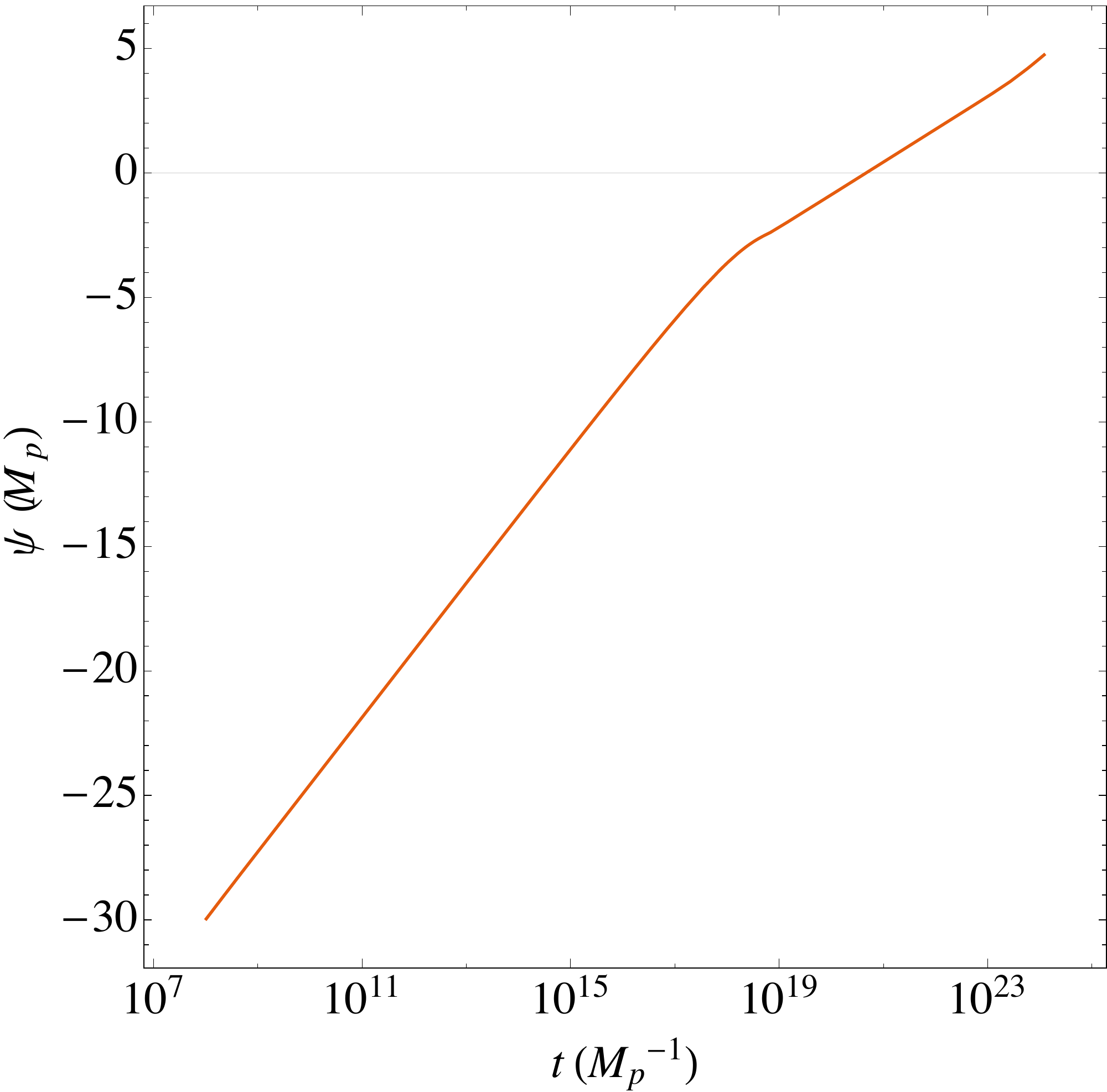}
\caption{Plots of $\varphi$ (left) and $\psi$ (right) vs $ t$ without friction. Notice that $\psi$ barely feels its minimum at $0$.}
\label{att4}
\end{figure}

But even with this ensured, there is generically an overshoot problem: $\psi$ typically acquires too much kinetic energy to prevent it climbing over the potential barrier and escaping towards infinity (see e.g., Figure \ref{att4}). Indeed, care must be taken not to overshoot the local maximum (which is not very high) and this is what we see happen for all the choices of initial conditions we explored. 

Past experience \cite{Albrecht:2001xt,Conlon:2008cj} shows there is a workaround to circumvent this if we can generate a significant component of radiation whose production and contribution to $H$ can enhance the friction experienced by rolling fields and thereby drain their kinetic energy. One way to do this would be to imagine that another field, $\chi$, has a $\varphi$-dependent mass that passes close to zero near a specific point, $\varphi = \varphi_f$, in field space. If so, then $\varphi$ begins to slow as it radiates $\chi$ particles when $\varphi \to \varphi_f$, generating a gas of radiation. Of course this gas becomes massive again once $\varphi$ leaves $\varphi_f$, a fact used to good effect in \cite{Kofman:2004yc} to dynamically trap $\varphi$ near $\varphi_f$. However the radiation can simply remain radiation if the $\chi$ fields can themselves decay very quickly into other very light particles whose masses are independent of $\varphi$. 

Within this type of scenario, the effects of particle production on field evolution can be captured by including also the energetics of the radiation and its production. The modified equations of motion are:
\bal
\ddot{\varphi} + \left(3 H + \Gamma\right) \dot{\varphi} + \frac{\partial W}{\partial \varphi}  &= 0\, , \nn\\
\ddot{\psi} + 3 H \dot{\psi} + \frac{\partial W}{\partial \psi}  &= 0\, , \,\, \text{and} \label{4D:EOM:fric}\\
\f {\dot{\varphi}^2}{2} + \f {\dot{\psi}^2}{2} + W + \rho &= 3\MP^2  H^2 \,, \nn
\eal
where $\Gamma = \Gamma(\varphi)$ is the decay rate for the production of radiation, which peaks near $\varphi = \varphi_f$. The conservation of energy requires the radiation energy density, $\rho$, to also satisfy
\bal
\dot \rho + 4 H \rho &= \Gamma \dot \varphi^2 \,.
\eal
With this kind of dynamics, it becomes feasible to trap $\psi$ if $\varphi_f$ is chosen to dump energy from the system into radiation at the right time (shortly before the radion reaches its local minimum). 

\begin{figure}[h]
\centering
\includegraphics[height=70mm]{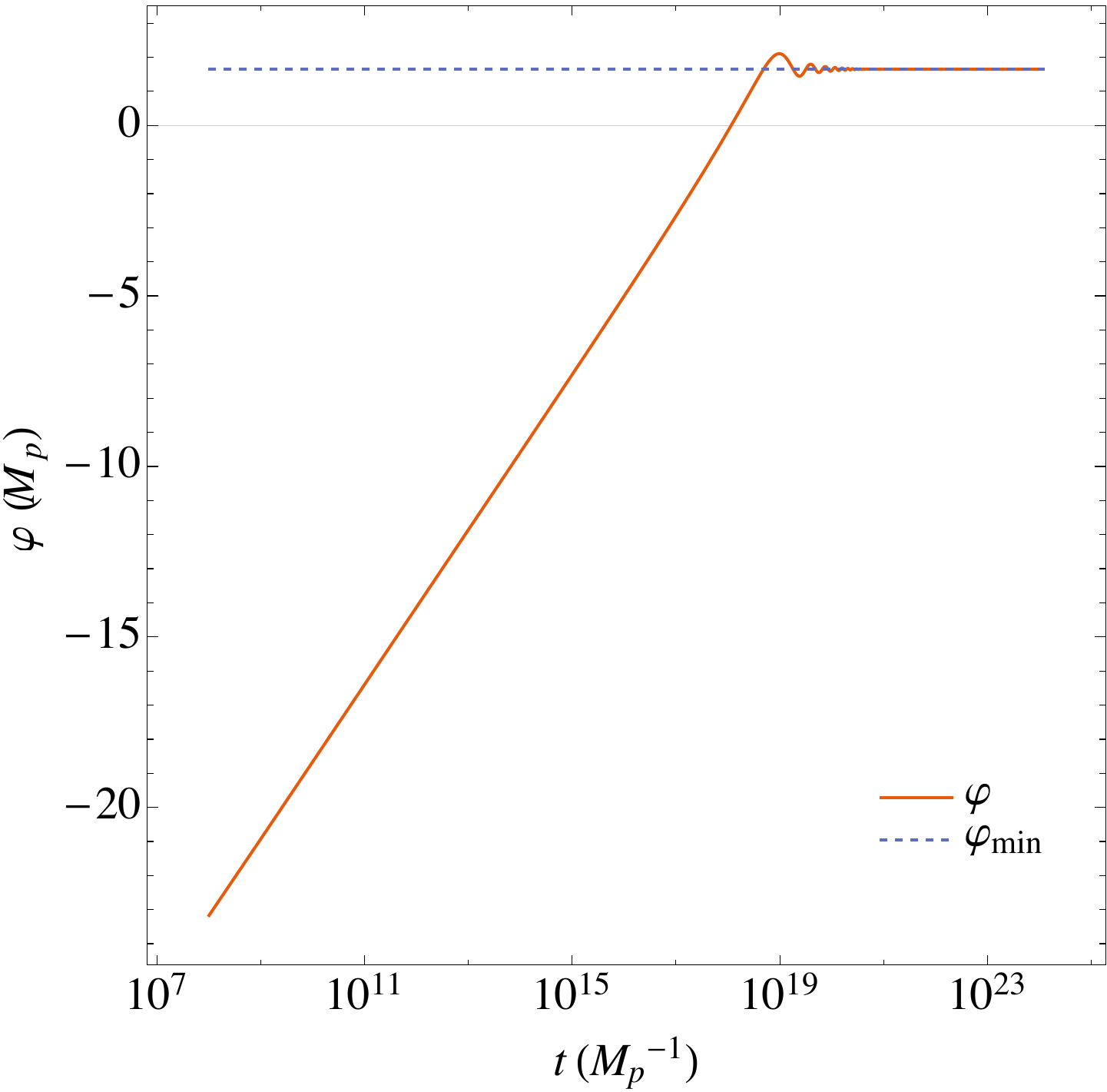}
\qquad
\includegraphics[height=70mm]{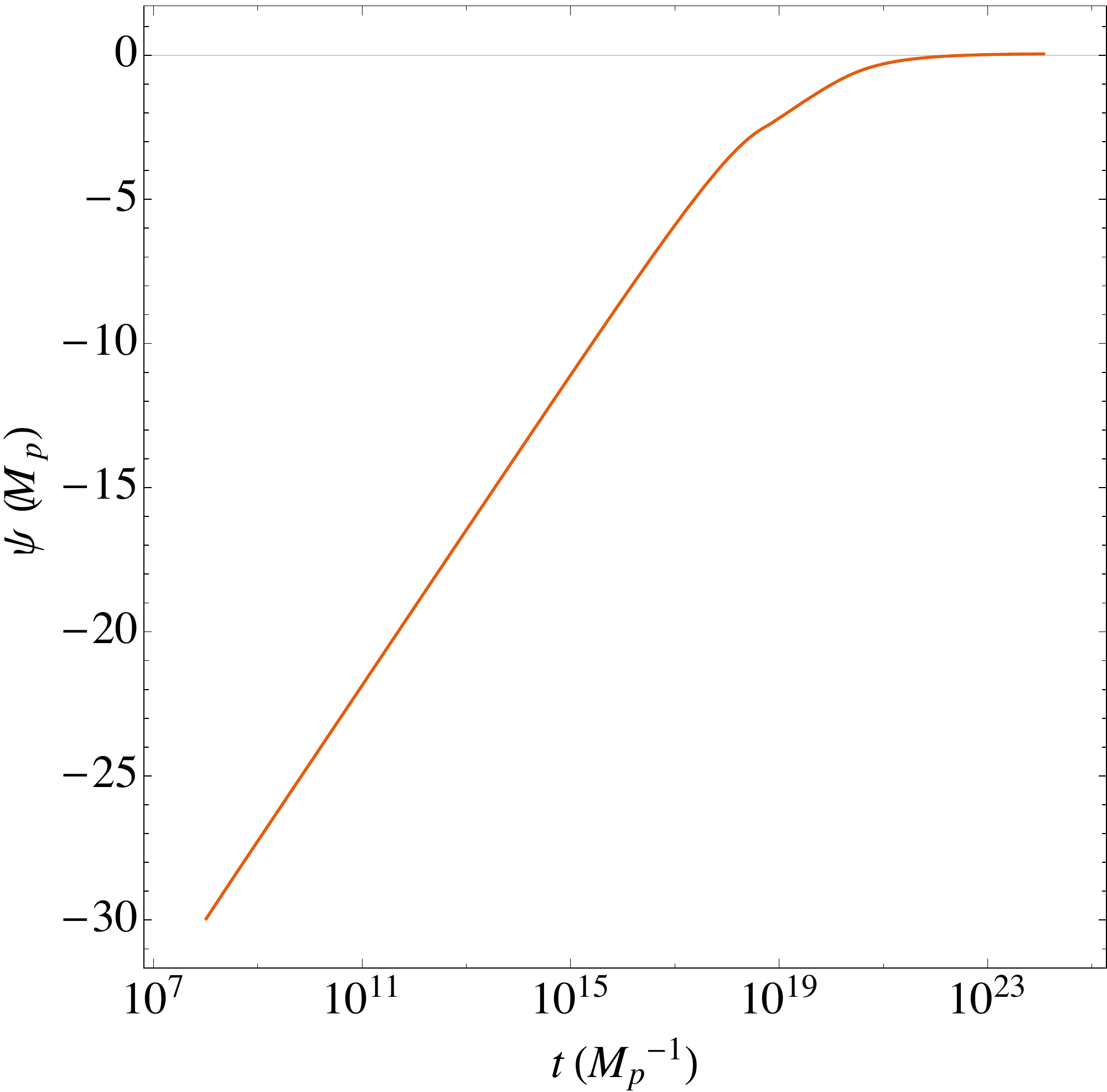}
\includegraphics[height=70mm]{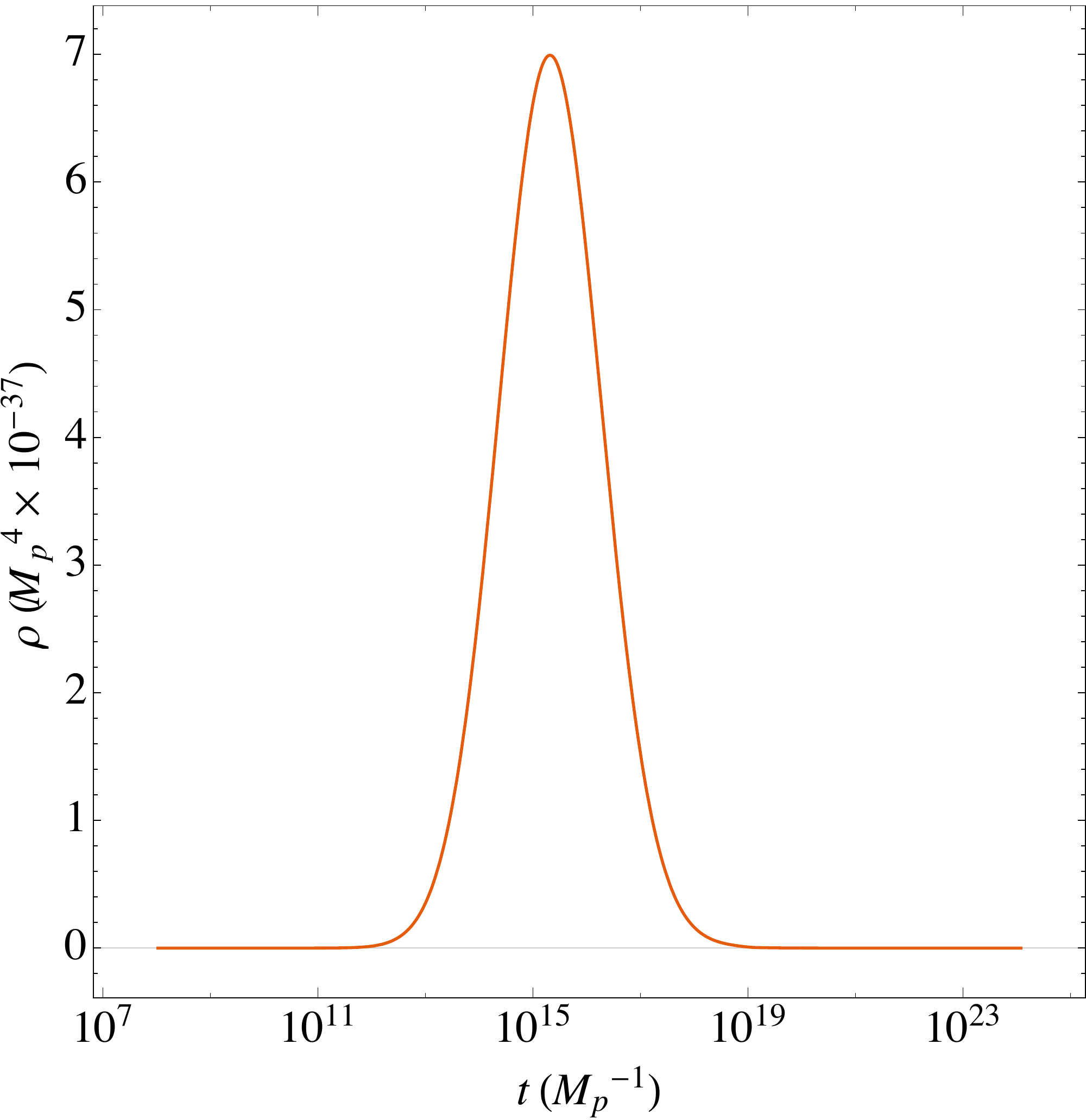}
\caption{Plots of $\varphi$ (top left) and $\psi$ (top right) vs $ t$ with a bout of friction (bottom) in $\varphi$ centred roughly at $ t \sim 10^{15}$. Having the radiation drain some energy from the system, $\psi$ is able to come in for a controlled landing.}
\label{att5}
\end{figure}

In our numerics, we assumed a Gaussian form for $\Gamma$, centred fairly close to the value of $\varphi$ for which $\psi \sim 0$. With this in place, we are able to trap both the inflaton and radion, as shown in the example illustrated in Figure \ref{att5}.

Even if the radion is trapped with such a contrivance, the resulting solution has one last drawback. Whereas one would expect from equation \eqref{pl:efolds} that there is a large enough phenomenologically allowed range for the extra dimensions to evolve, $b_f/b_0$, to allow a there to be 70 $e$-foldings for this solution. However, there is actually a stricter bound if we demand the radion be trapped at its minimum after the scaling solution ends. The stricter bound arises because the combined conditions that the flux can produce a minimum (i.e., \eqref{stab}) and that the stress-energy for the flux is sub-Planckian (i.e., \eqref{validity}), give us a relation between $b_0$ and $b_f = b_\star$:
\bal
b_0 \gg \left( \sqrt{4\pi} \, \f {b_f^3}{\MP}\right)^{1/4}.
\eal
As mentioned before, this power-law solution has $N_e = 2\ln(b_f/b_0)$, so we find:
\bal
N_e \ll \f 1 2\ln(b_f\MP) - \f 1 4 \ln(4\pi).
\eal
So, even if we allow the largest range possible for $b$ (with the final value, $b_f$, as large as the micron scale), we would still be left with $N_e \lesssim 35$.

\subsection{Slow-roll}

Our final example explores the numerics of configurations whose initial conditions ensure that the solution approximates the slow-roll scaling solution, discussed above, for part of the time. 

\begin{figure}[h]
\includegraphics[height=70mm]{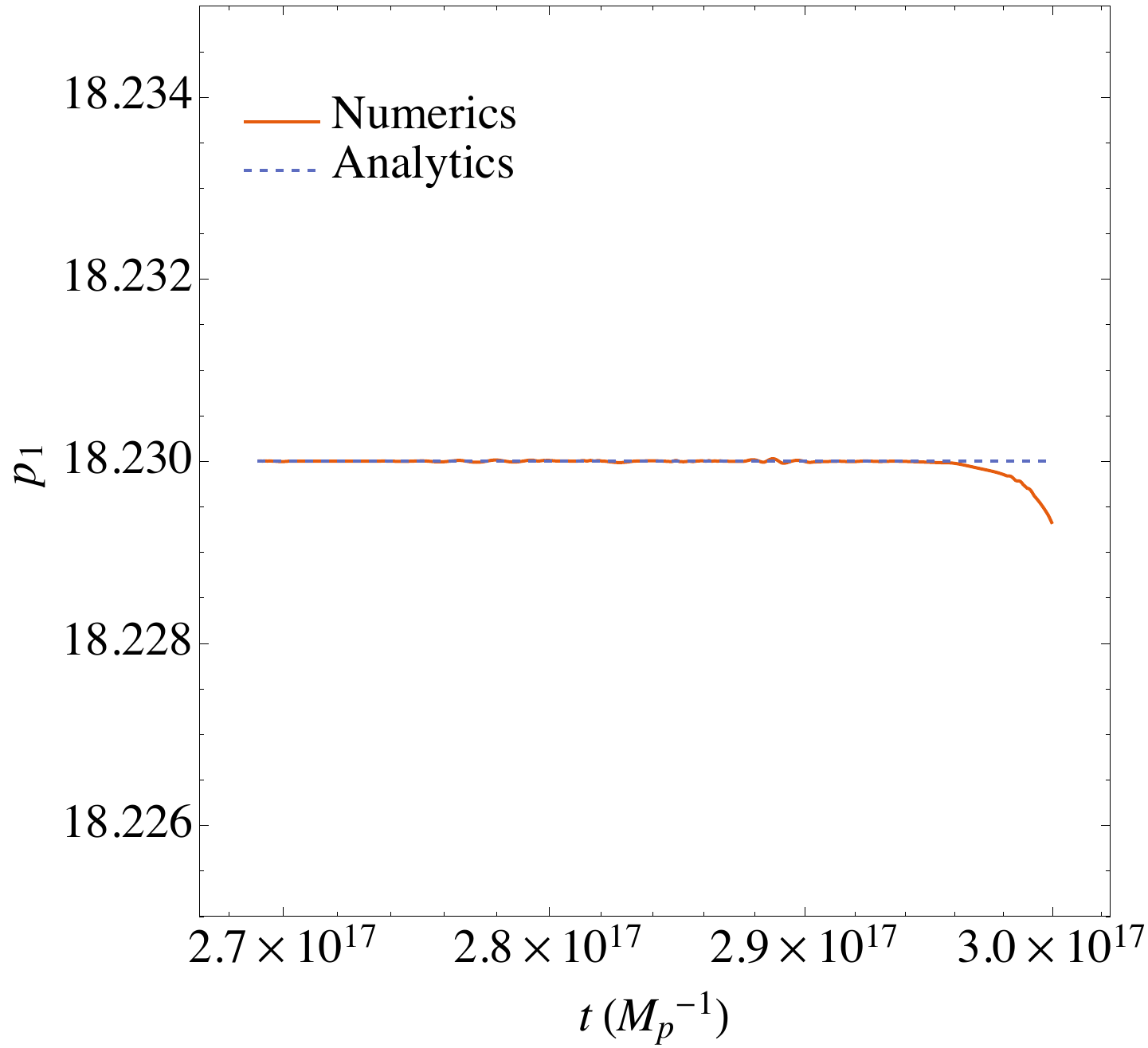}
\qquad
\includegraphics[height=70mm]{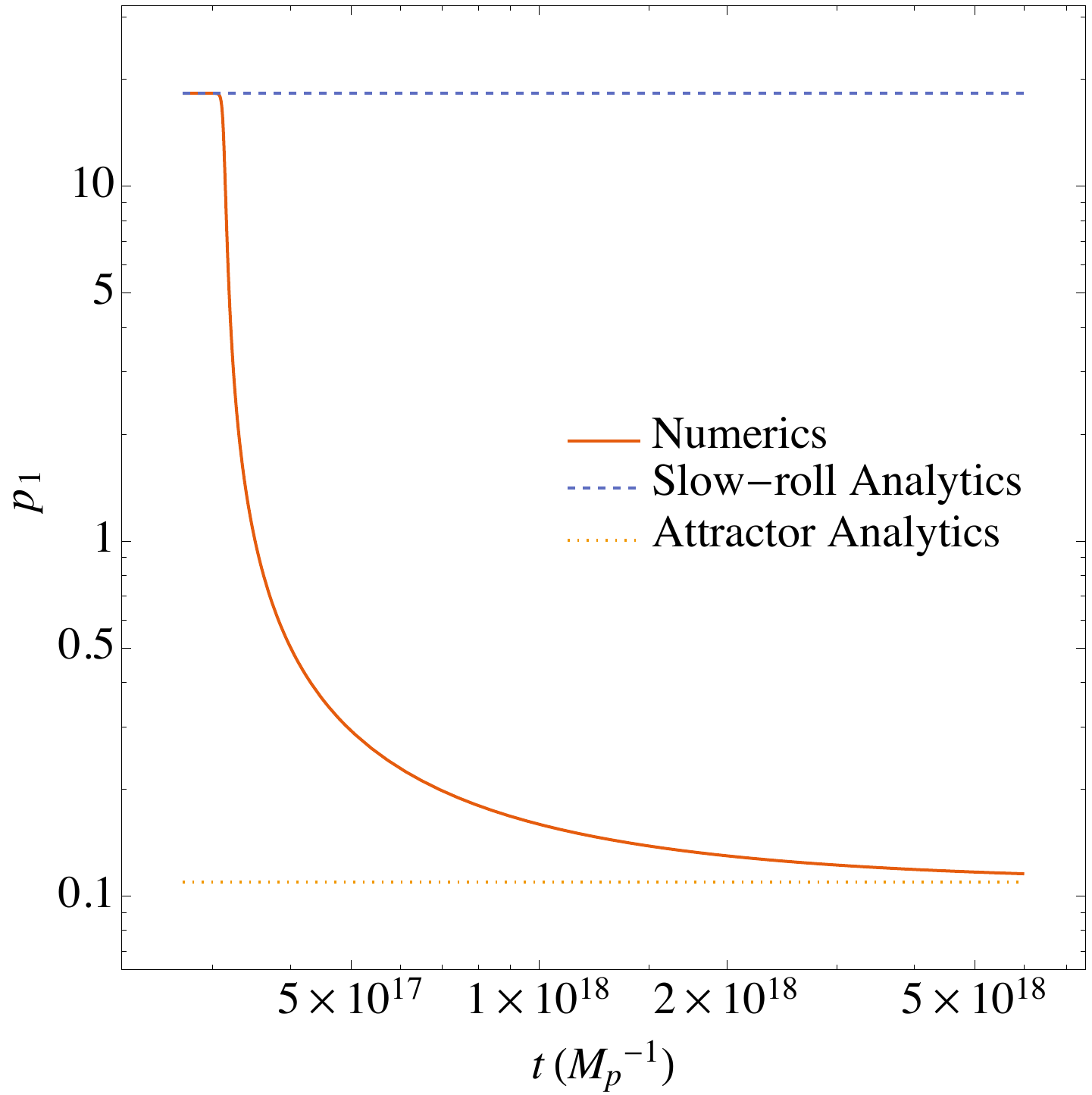} \\ \includegraphics[height=70mm]{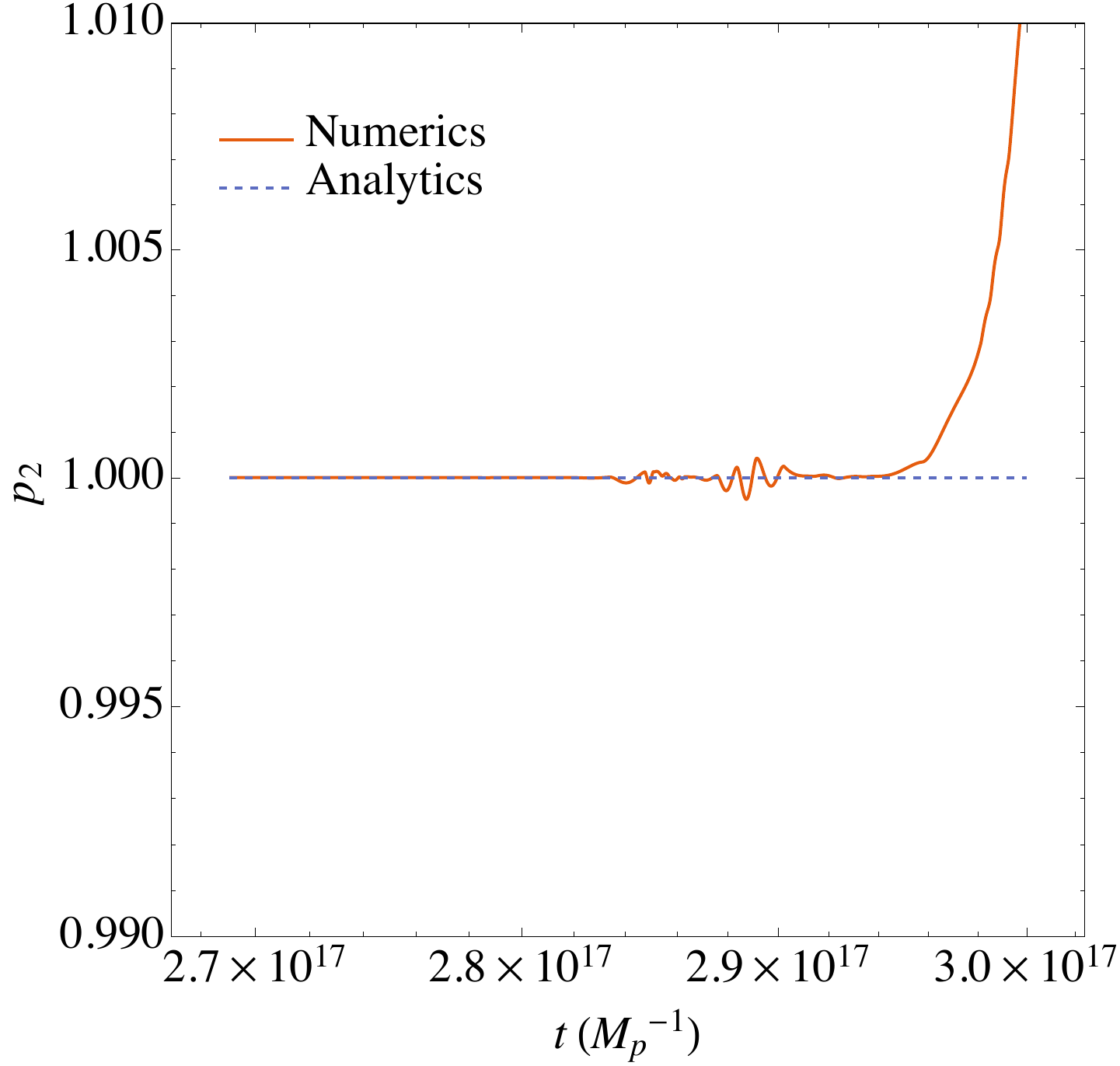}
\qquad
\includegraphics[height=70mm]{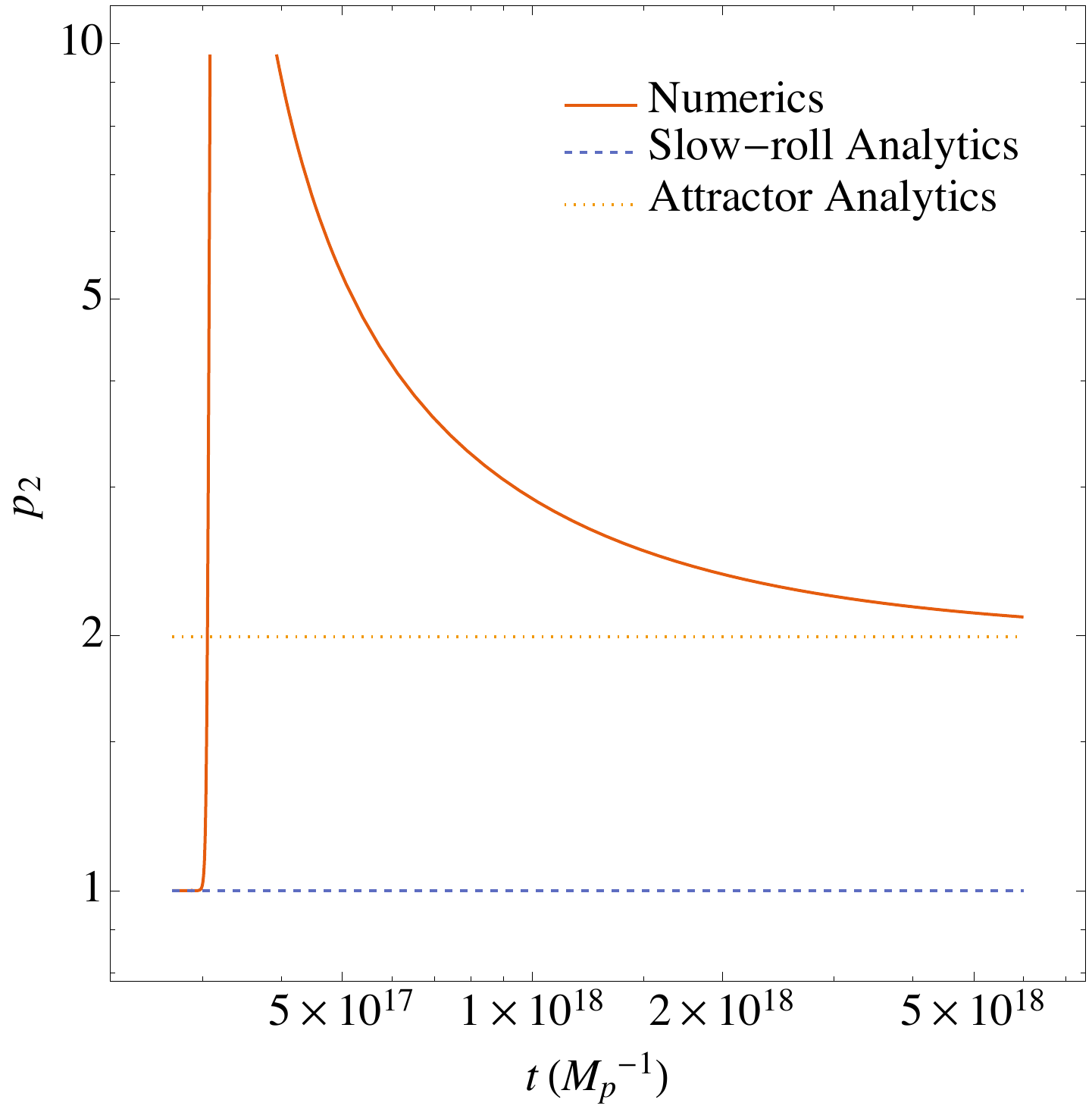}
\caption{Plots of $ t \dot \varphi$ (top) and $ t \dot \psi$ (bottom) vs $ t$ for the slow-roll scaling solution. On the left is an enlarged view of the slow-roll regime which highlights the agreement of our numerical solutions with the analytics. On the right, we see the transition from the unstable to the stable attractor solution.}
\label{sr1}
\end{figure}

\begin{figure}[h]
\centering
\includegraphics[width=0.8\textwidth]{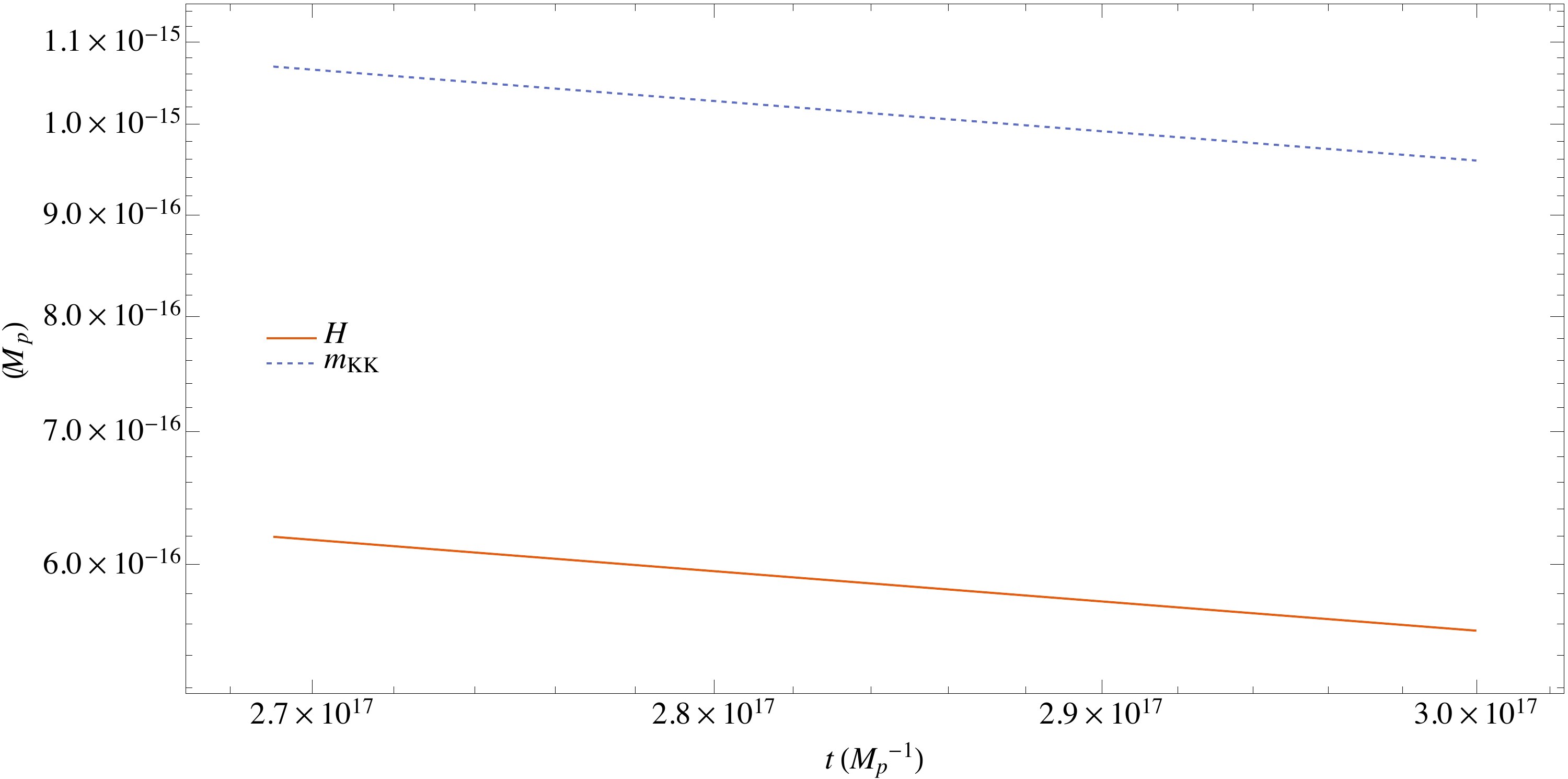}
\caption{Plot of $ H$ and $ m_{\kk}$ vs $ t$. Notice that we see the ratio $ H/ m_{\kk}$ is constant in time.}
\label{sr2}
\end{figure}

This solution is again characterized by the velocities of $\varphi$ and $\psi$ which are parametrized by $p_1$ and $p_2$. Figure \ref{sr1} shows a numerical solution with the expected behaviour, including the eventual departure from the scaling solution as expected given its unstable mode. Initial conditions were chosen to lie on the scaling solution, and we find the time spent there to be consistent with the numerical imprecision in these initial conditions. Figure \ref{sr1} also shows how the system tends towards the attractor solution once it leaves the slow-roll scaling regime. Figure \ref{sr2} plots the numerical evolution of the ratio $ H/ m_{\kk}$, and shows it is independent of time while in the slow-roll regime, as expected from the analytical evolution computed for the scaling solution. 

From the point of view of inflationary phenomenology the great virtue of this solution is the ability to arrange $\epsilon$ to be as small as we like (as well as the signature power-law result that $\eta = 0$). We exploit this freedom in the example illustrated to chose $\epsilon = 0.006$ (so that if 4D calculations of fluctuations were to apply we would expect $n_s \simeq 0.96$ and $r \simeq 0.096$). Our numerics agree with these predictions to within one part in $10^{8}$ and one part in $10^{5}$, for $ \epsilon$ and $ \eta$ respectively. 

The instability of the scaling solution is arguably a mixed benefit. On one hand it provides a simple way to end inflation since any mismatch between initial conditions and scaling solution is amplified even without changes to the scaling assumptions (such as the dominance of a single exponential in the scalar potential). This gives a simple way to dial in a large number of $e$-foldings, though at the expense of introducing specially chosen initial conditions.

On the other hand describing inflation through the slow-roll scaling solution seems inconsistent with simple trapping of the radion. This is because the defining condition that $\abs{W^{(\varphi)}} \sim \abs{W^{(c)}} \gg \abs{W^{(f)}}$ makes it impossible for this solution to end at a time when there is a flux-stabilized minimum for $\psi$. To see this remember that flux stabilizes the radion if the flux, inflaton potential, and extra-dimensional curvature stress-energies conspire at a certain point to have approximately the same magnitude (see equation \eqref{6D:static}). But given that the flux stress-energy scales as $b^{-4}$ while the curvature scales as $b^{-1}$, the flux term must dominate during inflation if they meet one another at inflation's end (as they would at the minimum, if it exists). Then at the start of inflation, the flux energy must have been significantly larger than that of the curvature, precluding the slow-roll scaling solution from starting. The exact time-dependence changes slightly in going from 6D to 4D, but the hierarchy remains the same, so we can see that, if we want a minimum to exist for $\psi$, it is impossible to start this solution in a regime that satisfies $\abs{W^{(\varphi)}} \sim \abs{W^{(c)}} \gg \abs{W^{(f)}}$. Consequently utilizing this solution for inflation would require stabilizing the extra-dimensional modulus with some other mechanism than the flux stabilization described above.

\section*{Acknowledgements}

We thank Ignatios Antoniadis and Sarah Shandera for useful discussions. CB thanks the Aspen Center for Physics, the Kavli Institute for Theoretical Physics (KITP), the NYU Center for Cosmology and Particle Physics (CCPP) and the TH department at CERN for their kind support while part of this work was done. This research was supported in part by funds from the Natural Sciences and Engineering Research Council (NSERC) of Canada and with funds from the Swiss National Science Foundation (SNSF). Research at the Perimeter Institute is supported in part by the Government of Canada through Industry Canada, and by the Province of Ontario through the Ministry of Research and Information (MRI). Work at KITP was supported in part by the National Science Foundation under Grant No. NSF PHY11-25915. Work at Aspen was supported in part by National Science Foundation Grant No. PHYS-1066293 and the hospitality of the Aspen Center for Physics.

\appendix

\section{Consistency of the truncation}
\label{App:Consistency}

Here, we demonstrate that the 4D equations of motion for homogeneous fields are fully equivalent to the nontrivial 6D equations of motion. This establishes consistency of the truncation because we have already checked that solving these 6D equations suffices to ensure a solution of the full 6D field equations, because the truncations required to obtain our {\it ansatz} are consistent with the equations of motion. This is as expected for an {\it ansatz} that is the most general consistent with a symmetry.

We first note that the changes of variables $ a = \hat a \,e^{\psi/2\MP}$ and $t' = \exd  t / \exd \hat  t = e^{\psi/2\MP}$ ensure
\be \label{HtoH}
 H = \f {\dot a} { a} = \left(\hat H + \f {\psi'}{2\MP} \right)e^{-\psi/2\MP} \qquad \hbox{while} \qquad
\cH = \frac{ \psi'}{2\MP} \,.
\ee
Substituting these into \eqref{4D:EOM} and using $\varphi = \sqrt{4\pi} \, \bnow \, \phi$ and the definition of $W(\psi, \varphi)$ gives:
\bea
&& -\f {\psi'}{2\MP} \, \phi' + \phi'' + 3\left(\f {\psi'}{2\MP} + \hat H \right)\phi' + \frac{\pd V}{\pd \phi} = 0\, , \nn \\
&& -\f { \psi'}{2\MP} \,\psi' + \psi'' + 3\left(\f { \psi'}{2\MP} + \hat H \right) \psi' - \f {4\pi \bnow^2}{\MP} \, V(\phi) + \f {2\MP}{\bnow^2} \, e^{-\psi/\MP} - \f {6\pi\mathfrak{f}^2}{\bnow^2\MP}  e^{-2\psi/\MP} = 0\, , \\
&& 4\pi\bnow^2 \f { (\phi')^2} 2 + \f {(\psi')^2} 2 + 4\pi \bnow^2 \, V(\phi) - \f {\MP^2}{\bnow^2}\,e^{-\psi/\MP} + \f {2\pi \mathfrak{f}^2}{\bnow^2} \, e^{-2\psi/\MP} = 3\MP^2 \left(\f {\psi'}{2\MP} + \hat H \right)^2 \,.\nn
\eea
Using $b = \bnow \exp(\psi/2\MP)$ to eliminate $\psi$ cleans these equations up to
\be
 \phi'' + \left(3 \hat H + 2\cH \right)\phi' + \frac{\pd V}{\pd \phi} = 0\, , \label{equiv1}
\ee
\be
\f {\psi''} {2\MP} + \left(3\hat H + 2\cH \right)\cH - \f {4\pi \bnow^2}{2\MP^2} V(\phi) + \f {1}{b^2} - \f {3\pi\bnow^2}{\MP^2} \f {\mathfrak{f}^2}{b^4} = 0\, , \label{almost}
\ee
and
\be
\f {4\pi\bnow^2}{\MP^2}\left\{ \f 1 2 \left[ \f {(\phi')^2} 2 + \f {\mathfrak{f}^2}{b^4}\right] + V(\phi)\right\} - \f {1}{b^2} = 3\hat H^2 + 6\hat H\cH + \cH^2 \,.\label{equiv2}
\ee
Using $\kappa^2 = 4\pi\bnow^2/\MP^2$ shows that \eqref{equiv1} is equivalent to the 6D inflaton equation of motion, \eqref{6EOM3}, while \eqref{equiv2} is equivalent to the 6D Friedmann equation (the first of eqs \eqref{EE}). Finally, using
\bal
\f {\psi''} {2\MP} &= \f { b''} b - \cH^2 
\eal
shows \eqref{almost} becomes
\bal
\f {b''} b + 3\hat H\cH + \cH^2 &= \f {4\pi\bnow^2}{\MP^2}\left( \f 1 2 V(\phi) + \f 3 4 \f {\mathfrak{f}^2}{b^4} \right) -  \f {1}{b^2}.
\eal
This is equivalent to a linear combination of the Einstein equations \eqref{EE}.  Solutions to the 4D equations therefore satisfy the homogeneous 6D equations and so (because the truncation is consistent) also exactly solve the full system of 6D classical field equations.

\section{Stability of scaling solutions}
\label{App:Stability}
In this Appendix, we study the stability of the scaling solutions introduced in the main text. We show in particular how one of these is stable and so an attractor for a broad basin of initial conditions, while others have unstable directions that control how long the scaling solution in question can dominate any particular numerical evolution.

We start by taking
\bal
\varphi &\rightarrow \varphi_\star + \delta\varphi, \nn\\
\psi &\rightarrow \psi_\star + \delta\psi, \,\, \text{and} \\
H &\rightarrow H_\star + \delta H,\nn
\eal
where (in this appendix only) $\varphi_\star, \psi_\star,$ and $H_\star$ are time-dependent solutions to the zeroth order equations of motion. The equations of motion \eqref{4D:EOM} to first order in these perturbations are:
\bal
\delta\ddot\varphi + 3\left(H_\star\delta\dot\varphi + \delta H \dot\varphi_\star \right) + \frac{ \lambda  U( t)}{\MP^2} \left(  \lambda \delta \varphi + \delta \psi \right) & = 0 
 \nn \\
\delta \ddot\psi + 3\left( H_\star \delta \dot\psi + \dot\psi_\star \delta H\right) + \left( \frac{ U( t)}{\MP^2} - \frac{4}{\bnow^2}\, e^{-2\psi_\star/\MP} \right) \delta \psi + \frac{ \lambda   U( t) }{\MP^2} \, \delta \varphi &=0 \label{eq:psip}\\
6 H_\star \MP^2 \, \delta H = \dot\varphi_\star \delta \dot\varphi + \dot\psi_\star \delta \dot\psi - \frac{  \lambda  U( t)}{\MP} \, \delta \varphi + \left(\frac{2 \MP}{\bnow^2} \, e^{-2\psi_\star/\MP} - \frac{ U( t)}{\MP} \right) \delta \psi 
\nn
\eal
where $ U( t) = 4 \pi \bnow^2 V_0 \exp[ -(  \lambda \varphi_\star + \psi_\star)/\MP]$ and we have assumed a potential of the form \eqref{scale:pot}. We then proceed to solve for $\delta H$ from the last of these and use it in the previous two equations of \pref{eq:psip}. Next, we substitute our power-law assumptions of the form \eqref{scale:psi}. This yields:
\bea
&&\delta \ddot\varphi + \left(3  \alpha + \frac{p_1^2}{2  \alpha } \right) \frac{\delta \dot\varphi }{ t} +\frac{  \lambda  U_0 }{\MP^2} \left(  \lambda - \frac{p_1}{2  \alpha } \right) T\, \delta \varphi +\left( \frac{p_1 p_2}{2  \alpha } \right) \frac{ \delta \dot\psi}{ t} \notag \\
&&\qquad\qquad\qquad \qquad + \left[\frac{ U_0}{\MP^2}\left(  \lambda - \frac{p_1}{2  \alpha } \right)T + \frac{p_1 }{ \alpha \bnow^2 } \, e^{-2 \psi_0/\MP} \left(\frac{ t}{ t_0}\right)^{-2p_2} \right] \delta \psi = 0  \label{eq:s:phi}\\
&&\delta \ddot\psi + \left(3  \alpha + \frac{p_2^2}{2  \alpha } \right)\frac{\delta \dot\psi}{ t} + \left[\frac{ U_0}{\MP^2} \left( 1 - \frac{p_2}{2  \alpha} \right) T + \frac{(p_2 - 4  \alpha ) }{ \alpha \bnow^2 } \, e^{-2 \psi_0/\MP} \left(\frac{ t}{ t_0}\right)^{-2p_2} \right] \delta \psi  \notag\\
&&\qquad \qquad\qquad \qquad +\left( \frac{p_1 p_2}{2  \alpha  } \right) \frac{\delta \dot\varphi}{ t} + \frac{ \lambda  U_0}{\MP^2} \left( 1 - \frac{p_2}{2  \alpha } \right) T \, \delta \varphi = 0 
\nn
\eea
where $ U_0$ is as previously defined in the main text and $T := \left({ t}/{ t_0}\right)^{- \lambda p_1 - p_2}$.

We now proceed to solve these equations in the attractor case, and slow-roll case to show that these cases are, indeed, attractors and unstable, respectively.

\subsection*{Attractor Solution}

Here, we again drop the terms that arise from perturbation in the flux terms (in equations \eqref{eq:s:phi}, these are the terms containing $1/\bnow^2$) and substitute our power-law solutions for this case, which are given by \eqref{att:alpha} and \eqref{att:coef}. This produces:
\bea \label{eq:pert2} 
&&\delta \ddot\varphi
+ \f{6+ \lambda^2 }{1+ \lambda^2 } \, \f{ \delta \dot\varphi}{  t}
+ \f{ \lambda^2(5- \lambda^2)}{(1+ \lambda^2 )^2}\, \f{ \delta  \varphi}{ t^2} 
+ \f{ \lambda }{1+ \lambda^2 }\, \f{ \delta \dot\psi}{ t} 
+ \f{ \lambda(5- \lambda^2)}{(1+ \lambda^2 )^2} \, \f{ \delta  \psi}{ t^2} = 0 \\
&& {\delta \ddot\psi} 
+ \f{7 }{1+ \lambda^2 }\, \f{ \delta \dot\psi}{ t} 
+ \f{5- \lambda^2}{(1+ \lambda^2 )^2} \, \f{ \delta  \psi}{ t^2}  
+ \f{ \lambda }{1+ \lambda^2 } \, \f{ \delta \dot\varphi}{ t}
+ \f{ \lambda(5- \lambda^2)}{(1+ \lambda^2 )^2} \, \f{ \delta  \varphi}{ t^2} = 0 \,. \nn
\eea
We model the perturbations as power-law solutions of the form
\bal \label{eq:pert:pl}
\delta \varphi = t^n, \qquad \hbox{and} \qquad  \delta \psi = A   t^m \, .
\eal
Substituting these into \eqref{eq:pert2} gives us two equations in $m$ and $n$:
\bea \label{eq:pertpoly}
&&\left[ ( \lambda^2 + 1)^2 n^2 + 5( \lambda^2 + 1) n + (5 -  \lambda^2) \lambda^2 \right] t^n \nn \\
&&\qquad\qquad\qquad
+ A \lambda \left[ ( \lambda^2 + 1)m + (5 -  \lambda ^2) \right]  t^m = 0 \, , \\
&& \lambda \left[ ( \lambda^2 + 1)n + (5 -  \lambda ^2) \right] t^n \nn \\
&&\qquad\qquad\qquad
+ A\left[ ( \lambda^2 + 1)^2 m^2 + 5( \lambda^2 + 1) m + (5 -  \lambda^2) \lambda^2 \right]  t^m  = 0 \, . \nn
\eea
If we assume $m\neq n$, then each polynomial in square brackets must vanish independently. However, since the polynomials for $n$ and $m$ are identical, they will have the same solution. We are left with $m=n$ and the following two equations which we then solve for $n$ and $A$:
\bea
( \lambda^2 + 1)^2 n^2 + ( \lambda^2 + 1)(5 + A  \lambda) n +  \lambda(5 -  \lambda^2)(A +  \lambda) &=& 0 \nn \\
A( \lambda^2 + 1)^2 n^2 + ( \lambda^2 + 1)(5A +   \lambda) n +  \lambda(5 -  \lambda^2)(1 + A  \lambda) &=& 0
\eea
This system has solutions:
\bea
n = \left(0, \, -1 \right), \qquad \hbox{while} \qquad A =\left( -  \lambda, \, \frac{1}{ \lambda}\right)
\eea
The other two linearly independent solutions are given by $\delta\varphi =  t^n,\,\, \delta\psi = 0$, and $\delta\varphi = 0,\,\, \delta\psi =  t^n$, where $n$ is given by $n = ( \lambda^2 - 5)/( \lambda^2 + 1)$.
Since $|  \lambda | < 1$, none of the solutions grow with time, hence this is a stable solution.

\subsection*{Slow-roll Solution}

As in the main text, we do not drop any term that arose from perturbations in the flux terms for this case. Substituting our power-law solutions for this case, which are given by \eqref{un:alpha} and \eqref{un:coef}, we produce:
\bea \label{eq:pert3}
&&{\delta \ddot\varphi} 
+ \f{3  \lambda^4 + 8 \lambda^2 +3}{2  \lambda^2 (1 +  \lambda^2) }\, \f{ \delta \dot\varphi}{  t}
+ \f{3+ \lambda^2}{2(1+ \lambda^2 )}\, \f{ \delta  \varphi}{  t^2} 
+ \f{ \lambda }{1+ \lambda^2 }\, \f{ \delta \dot\psi}{  t}
+ \f{3+ \lambda^2}{2 \lambda^3(1+ \lambda^2 )}\, \f{ \delta  \psi}{  t^2} = 0,  \nn \\
&&\delta\ddot{\psi} 
+ \f{5 \widetilde\lambda^4 + 6 \lambda^2 + 3 }{2  \lambda^2 (1 +  \lambda^2) }\f{ \delta \dot\psi}{  t}
+ \f{( \lambda^2+3)( \lambda^4+ \lambda^2-1)}{2 \lambda^4(1+ \lambda^2 )}\, \f{ \delta  \psi}{  t^2}  \\
&&\qquad\qquad\qquad\qquad\qquad\qquad\qquad
+ \f{ \lambda }{1+ \lambda^2 }\, \f{ \delta \dot\varphi}{  t}
+\f{3+ \lambda^2}{2  \lambda^3(1+ \lambda^2 )}\, \f{ \delta  \varphi}{  t^2}  = 0
\nn
\eea
Again, we model the perturbations as power-law solutions of the form given in \eqref{eq:pert:pl}.
We arrive at another two equations for $m$ and $n$:
\bea \label{eq:pertpoly2}
&& \lambda \left[ 2  \lambda ( \lambda^2 + 1)^2 n^2 + ( \lambda^4 + 6  \lambda^2  + 3) n + ( \lambda^2 + 3) \lambda^2 \right]  t^n \nn \\
&&\qquad\qquad\qquad 
+ A \lambda \left[ 2  \lambda^4 m + ( \lambda^2 + 3) \right]  t^m = 0 \, , \\
&& \lambda \left[ 2  \lambda^4 n + ( \lambda^2 + 3) \right]  t^n \nn \\
&&\qquad\qquad\qquad
+ A\left[ 2  \lambda^4 ( \lambda^2 + 1) m^2 +  \lambda^2(3  \lambda^4 + 4  \lambda^2 + 3) m +  \lambda^6 + 4  \lambda^4 - 3 \right]  t^m  = 0 \, . \nn
\eea
As before, these solutions are inconsistent if $m \neq n$, so, for the case where $ m = n$, we have another two equation in $n$ and $A$:
\bea
&&2  \lambda^3( \lambda ^2+1) n^2 
+ \lambda   (2 A  \lambda ^3+ \lambda ^4+6  \lambda ^2+3) n 
+ ( \lambda ^2+3) (A+ \lambda ^3)=0 \, , \nn \\
&&2 A  \lambda ^4 ( \lambda ^2+1)  n^2 
+  \lambda ^2 (A (3  \lambda ^4+4  \lambda ^2+3) +2  \lambda ^3)n  \\
&& \qquad \qquad \qquad \qquad \qquad
+ ( \lambda ^2+3) (A ( \lambda ^4+ \lambda ^2-1)+ \lambda ) =0 \, . \nn
\eea
This system has solutions:
\bea
&&n = \left(-1, \, -\f{ \lambda^2 + 3}{2 \lambda^2 }, \, -\f{3+  \lambda^2 \pm \sqrt{(11-7 \lambda^2)( \lambda^2 + 3)}}{4 \lambda^2} \right), \quad \hbox{while} \nn \\
&& A =\left(  \lambda, \,  \lambda, \, -\frac{1}{ \lambda}, \, -\frac{1}{ \lambda} \right) \, .
\eea
Since $|  \lambda^2| < 1$, we have one growing mode for $n$: hence, this solution is unstable.

  \bibliographystyle{JHEP}
  \bibliography{HDI}

\providecommand{\href}[2]{#2}\begingroup\raggedright\begin{thebibliography}{10}

\bibitem{Guth:1980zm}
A.~H. Guth, \emph{{The Inflationary Universe: A Possible Solution to the
  Horizon and Flatness Problems}},
  \href{http://dx.doi.org/10.1103/PhysRevD.23.347}{\emph{Phys. Rev.} {\bf D23}
  (1981) 347--356}.

\bibitem{Linde:1981mu}
A.~D. Linde, \emph{{A New Inflationary Universe Scenario: A Possible Solution
  of the Horizon, Flatness, Homogeneity, Isotropy and Primordial Monopole
  Problems}}, \href{http://dx.doi.org/10.1016/0370-2693(82)91219-9}{\emph{Phys.
  Lett.} {\bf B108} (1982) 389--393}.

\bibitem{Albrecht:1982wi}
A.~Albrecht and P.~J. Steinhardt, \emph{{Cosmology for Grand Unified Theories
  with Radiatively Induced Symmetry Breaking}},
  \href{http://dx.doi.org/10.1103/PhysRevLett.48.1220}{\emph{Phys. Rev. Lett.}
  {\bf 48} (1982) 1220--1223}.

\bibitem{Gasperini:1992em}
M.~Gasperini and G.~Veneziano, \emph{{Pre - big bang in string cosmology}},
  \href{http://dx.doi.org/10.1016/0927-6505(93)90017-8}{\emph{Astropart. Phys.}
  {\bf 1} (1993) 317--339}, [\href{http://arxiv.org/abs/hep-th/9211021}{{\tt
  hep-th/9211021}}].

\bibitem{Khoury:2001wf}
J.~Khoury, B.~A. Ovrut, P.~J. Steinhardt and N.~Turok, \emph{{The Ekpyrotic
  universe: Colliding branes and the origin of the hot big bang}},
  \href{http://dx.doi.org/10.1103/PhysRevD.64.123522}{\emph{Phys. Rev.} {\bf
  D64} (2001) 123522}, [\href{http://arxiv.org/abs/hep-th/0103239}{{\tt
  hep-th/0103239}}].

\bibitem{Ashtekar:2007em}
A.~Ashtekar, A.~Corichi and P.~Singh, \emph{{Robustness of key features of loop
  quantum cosmology}},
  \href{http://dx.doi.org/10.1103/PhysRevD.77.024046}{\emph{Phys. Rev.} {\bf
  D77} (2008) 024046}, [\href{http://arxiv.org/abs/0710.3565}{{\tt
  0710.3565}}].

\bibitem{Buchbinder:2007ad}
E.~I. Buchbinder, J.~Khoury and B.~A. Ovrut, \emph{{New Ekpyrotic cosmology}},
  \href{http://dx.doi.org/10.1103/PhysRevD.76.123503}{\emph{Phys. Rev.} {\bf
  D76} (2007) 123503}, [\href{http://arxiv.org/abs/hep-th/0702154}{{\tt
  hep-th/0702154}}].

\bibitem{Easson:2011zy}
D.~A. Easson, I.~Sawicki and A.~Vikman, \emph{{G-Bounce}},
  \href{http://dx.doi.org/10.1088/1475-7516/2011/11/021}{\emph{JCAP} {\bf 1111}
  (2011) 021}, [\href{http://arxiv.org/abs/1109.1047}{{\tt 1109.1047}}].

\bibitem{Kounnas:2011gz}
C.~Kounnas, H.~Partouche and N.~Toumbas, \emph{{S-brane to thermal non-singular
  string cosmology}},
  \href{http://dx.doi.org/10.1088/0264-9381/29/9/095014}{\emph{Class. Quant.
  Grav.} {\bf 29} (2012) 095014}, [\href{http://arxiv.org/abs/1111.5816}{{\tt
  1111.5816}}].

\bibitem{Brandenberger:2013zea}
R.~H. Brandenberger, C.~Kounnas, H.~Partouche, S.~P. Patil and N.~Toumbas,
  \emph{{Cosmological Perturbations Across an S-brane}},
  \href{http://dx.doi.org/10.1088/1475-7516/2014/03/015}{\emph{JCAP} {\bf 1403}
  (2014) 015}, [\href{http://arxiv.org/abs/1312.2524}{{\tt 1312.2524}}].

\bibitem{Baumann:2009ds}
D.~Baumann, \emph{{Inflation}},  in \emph{{Physics of the large and the small,
  TASI 09, proceedings of the Theoretical Advanced Study Institute in
  Elementary Particle Physics, Boulder, Colorado, USA, 1-26 June 2009}},
  pp.~523--686, 2011.
\newblock \href{http://arxiv.org/abs/0907.5424}{{\tt 0907.5424}}.
\newblock \href{http://dx.doi.org/10.1142/9789814327183_0010}{DOI}.

\bibitem{Novello:2008ra}
M.~Novello and S.~E.~P. Bergliaffa, \emph{{Bouncing Cosmologies}},
  \href{http://dx.doi.org/10.1016/j.physrep.2008.04.006}{\emph{Phys. Rept.}
  {\bf 463} (2008) 127--213}, [\href{http://arxiv.org/abs/0802.1634}{{\tt
  0802.1634}}].

\bibitem{Brandenberger:2016vhg}
R.~Brandenberger and P.~Peter, \emph{{Bouncing Cosmologies: Progress and
  Problems}},  \href{http://arxiv.org/abs/1603.05834}{{\tt 1603.05834}}.

\bibitem{Mukhanov:1981xt}
V.~F. Mukhanov and G.~V. Chibisov, \emph{{Quantum Fluctuation and Nonsingular
  Universe. (In Russian)}}, {\emph{JETP Lett.} {\bf 33} (1981) 532--535}.

\bibitem{Linde:1982uu}
A.~D. Linde, \emph{{Scalar Field Fluctuations in Expanding Universe and the New
  Inflationary Universe Scenario}},
  \href{http://dx.doi.org/10.1016/0370-2693(82)90293-3}{\emph{Phys. Lett.} {\bf
  B116} (1982) 335--339}.

\bibitem{Hawking:1982cz}
S.~W. Hawking, \emph{{The Development of Irregularities in a Single Bubble
  Inflationary Universe}},
  \href{http://dx.doi.org/10.1016/0370-2693(82)90373-2}{\emph{Phys. Lett.} {\bf
  B115} (1982) 295}.

\bibitem{Guth:1982ec}
A.~H. Guth and S.~Y. Pi, \emph{{Fluctuations in the New Inflationary
  Universe}}, \href{http://dx.doi.org/10.1103/PhysRevLett.49.1110}{\emph{Phys.
  Rev. Lett.} {\bf 49} (1982) 1110--1113}.

\bibitem{Bardeen:1983qw}
J.~M. Bardeen, P.~J. Steinhardt and M.~S. Turner, \emph{{Spontaneous Creation
  of Almost Scale - Free Density Perturbations in an Inflationary Universe}},
  \href{http://dx.doi.org/10.1103/PhysRevD.28.679}{\emph{Phys. Rev.} {\bf D28}
  (1983) 679}.

\bibitem{Ade:2015lrj}
{\scshape Planck} collaboration, P.~A.~R. Ade et~al., \emph{{Planck 2015
  results. XX. Constraints on inflation}},
  \href{http://arxiv.org/abs/1502.02114}{{\tt 1502.02114}}.

\bibitem{Brandenberger:1999sw}
R.~H. Brandenberger, \emph{{Inflationary cosmology: Progress and problems}},
  in \emph{{IPM School on Cosmology 1999: Large Scale Structure Formation
  Tehran, Iran, January 23-February 4, 1999}}, 1999.
\newblock \href{http://arxiv.org/abs/hep-ph/9910410}{{\tt hep-ph/9910410}}.

\bibitem{Burgess:2011fa}
C.~P. Burgess and L.~McAllister, \emph{{Challenges for String Cosmology}},
  \href{http://dx.doi.org/10.1088/0264-9381/28/20/204002}{\emph{Class. Quant.
  Grav.} {\bf 28} (2011) 204002}, [\href{http://arxiv.org/abs/1108.2660}{{\tt
  1108.2660}}].

\bibitem{Ijjas:2015hcc}
A.~Ijjas and P.~J. Steinhardt, \emph{{Implications of Planck2015 for
  inflationary, ekpyrotic and anamorphic bouncing cosmologies}},
  \href{http://dx.doi.org/10.1088/0264-9381/33/4/044001}{\emph{Class. Quant.
  Grav.} {\bf 33} (2016) 044001}, [\href{http://arxiv.org/abs/1512.09010}{{\tt
  1512.09010}}].

\bibitem{Cline:1999yq}
J.~M. Cline, \emph{{Cosmological expansion in the Randall-Sundrum warped
  compactification}},  in \emph{{Proceedings, 3rd International Conference on
  Particle Physics and the Early Universe (COSMO 1999)}}, pp.~472--479, 2000.
\newblock \href{http://arxiv.org/abs/hep-ph/0001285}{{\tt hep-ph/0001285}}.
\newblock \href{http://dx.doi.org/10.1142/9789812792129_0072}{DOI}.

\bibitem{Flanagan:1999cu}
E.~E. Flanagan, S.~H.~H. Tye and I.~Wasserman, \emph{{Cosmological expansion in
  the Randall-Sundrum brane world scenario}},
  \href{http://dx.doi.org/10.1103/PhysRevD.62.044039}{\emph{Phys. Rev.} {\bf
  D62} (2000) 044039}, [\href{http://arxiv.org/abs/hep-ph/9910498}{{\tt
  hep-ph/9910498}}].

\bibitem{Kim:2000hi}
H.~B. Kim, \emph{{Cosmology of Randall-Sundrum models with an extra dimension
  stabilized by balancing bulk matter}},
  \href{http://dx.doi.org/10.1016/S0370-2693(00)00251-3}{\emph{Phys. Lett.}
  {\bf B478} (2000) 285--293}, [\href{http://arxiv.org/abs/hep-th/0001209}{{\tt
  hep-th/0001209}}].

\bibitem{Lesgourgues:2000tj}
J.~Lesgourgues, S.~Pastor, M.~Peloso and L.~Sorbo, \emph{{Cosmology of the
  Randall-Sundrum model after dilaton stabilization}},
  \href{http://dx.doi.org/10.1016/S0370-2693(00)00943-6}{\emph{Phys. Lett.}
  {\bf B489} (2000) 411}, [\href{http://arxiv.org/abs/hep-ph/0004086}{{\tt
  hep-ph/0004086}}].

\bibitem{Randall:1999ee}
L.~Randall and R.~Sundrum, \emph{{A Large mass hierarchy from a small extra
  dimension}}, \href{http://dx.doi.org/10.1103/PhysRevLett.83.3370}{\emph{Phys.
  Rev. Lett.} {\bf 83} (1999) 3370--3373},
  [\href{http://arxiv.org/abs/hep-ph/9905221}{{\tt hep-ph/9905221}}].

\bibitem{Linde:2005dd}
A.~D. Linde, \emph{{Inflation and string cosmology}},
  \href{http://dx.doi.org/10.1143/PTPS.163.295}{\emph{Prog. Theor. Phys.
  Suppl.} {\bf 163} (2006) 295--322},
  [\href{http://arxiv.org/abs/hep-th/0503195}{{\tt hep-th/0503195}}].

\bibitem{Burgess:2007pz}
C.~P. Burgess, \emph{{Lectures on Cosmic Inflation and its Potential Stringy
  Realizations}},
  \href{http://dx.doi.org/10.1088/0264-9381/24/21/S04}{\emph{Class. Quant.
  Grav.} {\bf 24} (2007) S795}, [\href{http://arxiv.org/abs/0708.2865}{{\tt
  0708.2865}}].

\bibitem{Baumann:2009ni}
D.~Baumann and L.~McAllister, \emph{{Advances in Inflation in String Theory}},
  \href{http://dx.doi.org/10.1146/annurev.nucl.010909.083524}{\emph{Ann. Rev.
  Nucl. Part. Sci.} {\bf 59} (2009) 67--94},
  [\href{http://arxiv.org/abs/0901.0265}{{\tt 0901.0265}}].

\bibitem{Cicoli:2011zz}
M.~Cicoli and F.~Quevedo, \emph{{String moduli inflation: An overview}},
  \href{http://dx.doi.org/10.1088/0264-9381/28/20/204001}{\emph{Class. Quant.
  Grav.} {\bf 28} (2011) 204001}, [\href{http://arxiv.org/abs/1108.2659}{{\tt
  1108.2659}}].

\bibitem{Burgess:2013sla}
C.~P. Burgess, M.~Cicoli and F.~Quevedo, \emph{{String Inflation After Planck
  2013}}, \href{http://dx.doi.org/10.1088/1475-7516/2013/11/003}{\emph{JCAP}
  {\bf 1311} (2013) 003}, [\href{http://arxiv.org/abs/1306.3512}{{\tt
  1306.3512}}].

\bibitem{Silverstein:2015mll}
E.~Silverstein, \emph{{Inflation in string theory confronts data/Les mod\`eles
  d'inflation en th\'eorie des cordes face aux observations}},
  \href{http://arxiv.org/abs/1512.02089}{{\tt 1512.02089}}.

\bibitem{vanNierop:2011di}
L.~van Nierop and C.~P. Burgess, \emph{{Sculpting the Extra Dimensions:
  Inflation from Codimension-2 Brane Back-reaction}},
  \href{http://dx.doi.org/10.1088/1475-7516/2012/04/037}{\emph{JCAP} {\bf 1204}
  (2012) 037}, [\href{http://arxiv.org/abs/1108.2553}{{\tt 1108.2553}}].

\bibitem{Arkani-Hamed:2015bza}
N.~Arkani-Hamed and J.~Maldacena, \emph{{Cosmological Collider Physics}},
  \href{http://arxiv.org/abs/1503.08043}{{\tt 1503.08043}}.

\bibitem{Higuchi:1986py}
A.~Higuchi, \emph{{Forbidden Mass Range for Spin-2 Field Theory in De Sitter
  Space-time}},
  \href{http://dx.doi.org/10.1016/0550-3213(87)90691-2}{\emph{Nucl. Phys.} {\bf
  B282} (1987) 397}.

\bibitem{Abbott:1984fp}
L.~F. Abbott and M.~B. Wise, \emph{{Constraints on Generalized Inflationary
  Cosmologies}},
  \href{http://dx.doi.org/10.1016/0550-3213(84)90329-8}{\emph{Nucl. Phys.} {\bf
  B244} (1984) 541--548}.

\bibitem{Yokoyama:1987an}
J.~Yokoyama and K.-i. Maeda, \emph{{On the Dynamics of the Power Law Inflation
  Due to an Exponential Potential}},
  \href{http://dx.doi.org/10.1016/0370-2693(88)90880-5}{\emph{Phys. Lett.} {\bf
  B207} (1988) 31}.

\bibitem{Liddle:1988tb}
A.~R. Liddle, \emph{{Power Law Inflation With Exponential Potentials}},
  \href{http://dx.doi.org/10.1016/0370-2693(89)90776-4}{\emph{Phys. Lett.} {\bf
  B220} (1989) 502}.

\bibitem{PhysRevD.32.1316}
F.~Lucchin and S.~Matarrese, \emph{Power-law inflation},
  \href{http://dx.doi.org/10.1103/PhysRevD.32.1316}{\emph{Phys. Rev. D} {\bf
  32} (Sep, 1985) 1316--1322}.

\bibitem{La:1989za}
D.~La and P.~J. Steinhardt, \emph{{Extended Inflationary Cosmology}},
  \href{http://dx.doi.org/10.1103/PhysRevLett.62.376}{\emph{Phys. Rev. Lett.}
  {\bf 62} (1989) 376}.

\bibitem{Weinberg:1989mp}
E.~J. Weinberg, \emph{{Some Problems with Extended Inflation}},
  \href{http://dx.doi.org/10.1103/PhysRevD.40.3950}{\emph{Phys. Rev.} {\bf D40}
  (1989) 3950}.

\bibitem{Liddle:1992wi}
A.~R. Liddle and D.~H. Lyth, \emph{{COBE, gravitational waves, inflation and
  extended inflation}},
  \href{http://dx.doi.org/10.1016/0370-2693(92)91393-N}{\emph{Phys. Lett.} {\bf
  B291} (1992) 391--398}, [\href{http://arxiv.org/abs/astro-ph/9208007}{{\tt
  astro-ph/9208007}}].

\bibitem{Copeland:1997et}
E.~J. Copeland, A.~R. Liddle and D.~Wands, \emph{{Exponential potentials and
  cosmological scaling solutions}},
  \href{http://dx.doi.org/10.1103/PhysRevD.57.4686}{\emph{Phys. Rev.} {\bf D57}
  (1998) 4686--4690}, [\href{http://arxiv.org/abs/gr-qc/9711068}{{\tt
  gr-qc/9711068}}].

\bibitem{Array:2015xqh}
{\scshape BICEP2, Keck Array} collaboration, P.~A.~R. Ade et~al.,
  \emph{{Improved Constraints on Cosmology and Foregrounds from BICEP2 and Keck
  Array Cosmic Microwave Background Data with Inclusion of 95 GHz Band}},
  \href{http://dx.doi.org/10.1103/PhysRevLett.116.031302}{\emph{Phys. Rev.
  Lett.} {\bf 116} (2016) 031302}, [\href{http://arxiv.org/abs/1510.09217}{{\tt
  1510.09217}}].

\bibitem{Freund:1980xh}
P.~G.~O. Freund and M.~A. Rubin, \emph{{Dynamics of Dimensional Reduction}},
  \href{http://dx.doi.org/10.1016/0370-2693(80)90590-0}{\emph{Phys. Lett.} {\bf
  B97} (1980) 233--235}.

\bibitem{Giddings:2001yu}
S.~B. Giddings, S.~Kachru and J.~Polchinski, \emph{{Hierarchies from fluxes in
  string compactifications}},
  \href{http://dx.doi.org/10.1103/PhysRevD.66.106006}{\emph{Phys. Rev.} {\bf
  D66} (2002) 106006}, [\href{http://arxiv.org/abs/hep-th/0105097}{{\tt
  hep-th/0105097}}].

\bibitem{Dasgupta:1999ss}
K.~Dasgupta, G.~Rajesh and S.~Sethi, \emph{{M theory, orientifolds and G -
  flux}}, \href{http://dx.doi.org/10.1088/1126-6708/1999/08/023}{\emph{JHEP}
  {\bf 08} (1999) 023}, [\href{http://arxiv.org/abs/hep-th/9908088}{{\tt
  hep-th/9908088}}].

\bibitem{Kachru:2003aw}
S.~Kachru, R.~Kallosh, A.~D. Linde and S.~P. Trivedi, \emph{{De Sitter vacua in
  string theory}},
  \href{http://dx.doi.org/10.1103/PhysRevD.68.046005}{\emph{Phys. Rev.} {\bf
  D68} (2003) 046005}, [\href{http://arxiv.org/abs/hep-th/0301240}{{\tt
  hep-th/0301240}}].

\bibitem{Kachru:2003sx}
S.~Kachru, R.~Kallosh, A.~D. Linde, J.~M. Maldacena, L.~P. McAllister and S.~P.
  Trivedi, \emph{{Towards inflation in string theory}},
  \href{http://dx.doi.org/10.1088/1475-7516/2003/10/013}{\emph{JCAP} {\bf 0310}
  (2003) 013}, [\href{http://arxiv.org/abs/hep-th/0308055}{{\tt
  hep-th/0308055}}].

\bibitem{BlancoPillado:2004ns}
J.~J. Blanco-Pillado, C.~P. Burgess, J.~M. Cline, C.~Escoda, M.~Gomez-Reino,
  R.~Kallosh et~al., \emph{{Racetrack inflation}},
  \href{http://dx.doi.org/10.1088/1126-6708/2004/11/063}{\emph{JHEP} {\bf 11}
  (2004) 063}, [\href{http://arxiv.org/abs/hep-th/0406230}{{\tt
  hep-th/0406230}}].

\bibitem{Freund:1982pg}
P.~G.~O. Freund, \emph{{Kaluza-Klein Cosmologies}},
  \href{http://dx.doi.org/10.1016/0550-3213(82)90106-7}{\emph{Nucl. Phys.} {\bf
  B209} (1982) 146}.

\bibitem{RandjbarDaemi:1983jz}
S.~Randjbar-Daemi, A.~Salam and J.~A. Strathdee, \emph{{On Kaluza-Klein
  Cosmology}},
  \href{http://dx.doi.org/10.1016/0370-2693(84)90300-9}{\emph{Phys. Lett.} {\bf
  B135} (1984) 388--392}.

\bibitem{Shafi:1984ha}
Q.~Shafi and C.~Wetterich, \emph{{Inflation With Higher Dimensional Gravity}},
  \href{http://dx.doi.org/10.1016/0370-2693(85)91137-2}{\emph{Phys. Lett.} {\bf
  B152} (1985) 51}.

\bibitem{Abbott:1984ba}
R.~B. Abbott, S.~M. Barr and S.~D. Ellis, \emph{{Kaluza-Klein Cosmologies and
  Inflation}}, \href{http://dx.doi.org/10.1103/PhysRevD.30.720}{\emph{Phys.
  Rev.} {\bf D30} (1984) 720}.

\bibitem{Sahdev:1988fp}
D.~Sahdev, \emph{{Towards a Realistic Kaluza-Klein Cosmology}},
  \href{http://dx.doi.org/10.1016/0370-2693(84)90220-X}{\emph{Phys. Lett.} {\bf
  B137} (1984) 155--159}.

\bibitem{Kaloper:2000jb}
N.~Kaloper, J.~March-Russell, G.~D. Starkman and M.~Trodden, \emph{{Compact
  hyperbolic extra dimensions: Branes, Kaluza-Klein modes and cosmology}},
  \href{http://dx.doi.org/10.1103/PhysRevLett.85.928}{\emph{Phys. Rev. Lett.}
  {\bf 85} (2000) 928--931}, [\href{http://arxiv.org/abs/hep-ph/0002001}{{\tt
  hep-ph/0002001}}].

\bibitem{Okada:1984cv}
Y.~Okada, \emph{{Inflation in {Kaluza-Klein} Cosmology}},
  \href{http://dx.doi.org/10.1016/0370-2693(85)90148-0}{\emph{Phys. Lett.} {\bf
  B150} (1985) 103--106}.

\bibitem{Maeda:1984gq}
K.-i. Maeda and H.~Nishino, \emph{{Cosmological Solutions in $D=6$, $N=2$
  {Kaluza-Klein} Supergravity: Friedmann Universe Without Fine Tuning}},
  \href{http://dx.doi.org/10.1016/0370-2693(85)90409-5}{\emph{Phys. Lett.} {\bf
  B154} (1985) 358--362}.

\bibitem{WETTERICH1985309}
C.~Wetterich, \emph{Kaluza-klein cosmology and the inflationary universe},
  \href{http://dx.doi.org/http://dx.doi.org/10.1016/0550-3213(85)90445-6}{\emph{Nuclear
  Physics B} {\bf 252} (1985) 309 -- 320}.

\bibitem{Gunther:2000jj}
U.~Gunther and A.~Zhuk, \emph{{Stabilization of internal spaces in
  multidimensional cosmology}},
  \href{http://dx.doi.org/10.1103/PhysRevD.61.124001}{\emph{Phys. Rev.} {\bf
  D61} (2000) 124001}, [\href{http://arxiv.org/abs/hep-ph/0002009}{{\tt
  hep-ph/0002009}}].

\bibitem{Gunther:2000yb}
U.~Gunther and A.~Zhuk, \emph{{A Note on dynamical stabilization of internal
  spaces in multidimensional cosmology}},
  \href{http://dx.doi.org/10.1088/0264-9381/18/8/303}{\emph{Class. Quant.
  Grav.} {\bf 18} (2001) 1441--1460},
  [\href{http://arxiv.org/abs/hep-ph/0006283}{{\tt hep-ph/0006283}}].

\bibitem{BlancoPillado:2011me}
J.~J. Blanco-Pillado, H.~S. Ramadhan and B.~Shlaer, \emph{{Bubbles from
  Nothing}}, \href{http://dx.doi.org/10.1088/1475-7516/2012/01/045}{\emph{JCAP}
  {\bf 1201} (2012) 045}, [\href{http://arxiv.org/abs/1104.5229}{{\tt
  1104.5229}}].

\bibitem{Grana:2005jc}
M.~Grana, \emph{{Flux compactifications in string theory: A Comprehensive
  review}}, \href{http://dx.doi.org/10.1016/j.physrep.2005.10.008}{\emph{Phys.
  Rept.} {\bf 423} (2006) 91--158},
  [\href{http://arxiv.org/abs/hep-th/0509003}{{\tt hep-th/0509003}}].

\bibitem{RandjbarDaemi:1982hi}
S.~Randjbar-Daemi, A.~Salam and J.~A. Strathdee, \emph{{Spontaneous
  Compactification in Six-Dimensional Einstein-Maxwell Theory}},
  \href{http://dx.doi.org/10.1016/0550-3213(83)90247-X}{\emph{Nucl. Phys.} {\bf
  B214} (1983) 491--512}.

\bibitem{Salam:1984cj}
A.~Salam and E.~Sezgin, \emph{{Chiral Compactification on Minkowski x S**2 of
  N=2 Einstein-Maxwell Supergravity in Six-Dimensions}},
  \href{http://dx.doi.org/10.1016/0370-2693(84)90589-6}{\emph{Phys. Lett.} {\bf
  B147} (1984) 47}.

\bibitem{Maeda:1985bq}
K.-i. Maeda, \emph{{STABILITY AND ATTRACTOR IN KALUZA-KLEIN COSMOLOGY. 1.}},
  \href{http://dx.doi.org/10.1088/0264-9381/3/2/017}{\emph{Class. Quant. Grav.}
  {\bf 3} (1986) 233}.

\bibitem{BlancoPillado:2009di}
J.~J. Blanco-Pillado, D.~Schwartz-Perlov and A.~Vilenkin, \emph{{Quantum
  Tunneling in Flux Compactifications}},
  \href{http://dx.doi.org/10.1088/1475-7516/2009/12/006}{\emph{JCAP} {\bf 0912}
  (2009) 006}, [\href{http://arxiv.org/abs/0904.3106}{{\tt 0904.3106}}].

\bibitem{BlancoPillado:2009mi}
J.~J. Blanco-Pillado, D.~Schwartz-Perlov and A.~Vilenkin,
  \emph{{Transdimensional Tunneling in the Multiverse}},
  \href{http://dx.doi.org/10.1088/1475-7516/2010/05/005}{\emph{JCAP} {\bf 1005}
  (2010) 005}, [\href{http://arxiv.org/abs/0912.4082}{{\tt 0912.4082}}].

\bibitem{Brown:2010bc}
A.~R. Brown and A.~Dahlen, \emph{{Small Steps and Giant Leaps in the
  Landscape}}, \href{http://dx.doi.org/10.1103/PhysRevD.82.083519}{\emph{Phys.
  Rev.} {\bf D82} (2010) 083519}, [\href{http://arxiv.org/abs/1004.3994}{{\tt
  1004.3994}}].

\bibitem{Brown:2010mf}
A.~R. Brown and A.~Dahlen, \emph{{Bubbles of Nothing and the Fastest Decay in
  the Landscape}},
  \href{http://dx.doi.org/10.1103/PhysRevD.84.043518}{\emph{Phys. Rev.} {\bf
  D84} (2011) 043518}, [\href{http://arxiv.org/abs/1010.5240}{{\tt
  1010.5240}}].

\bibitem{Brown:2010mg}
A.~R. Brown and A.~Dahlen, \emph{{Giant Leaps and Minimal Branes in
  Multi-Dimensional Flux Landscapes}},
  \href{http://dx.doi.org/10.1103/PhysRevD.84.023513}{\emph{Phys. Rev.} {\bf
  D84} (2011) 023513}, [\href{http://arxiv.org/abs/1010.5241}{{\tt
  1010.5241}}].

\bibitem{Brown:2011gt}
A.~R. Brown and A.~Dahlen, \emph{{On 'nothing' as an infinitely negatively
  curved spacetime}},
  \href{http://dx.doi.org/10.1103/PhysRevD.85.104026}{\emph{Phys. Rev.} {\bf
  D85} (2012) 104026}, [\href{http://arxiv.org/abs/1111.0301}{{\tt
  1111.0301}}].

\bibitem{Burgess:2003jk}
C.~P. Burgess, \emph{{Quantum gravity in everyday life: General relativity as
  an effective field theory}},
  \href{http://dx.doi.org/10.12942/lrr-2004-5}{\emph{Living Rev. Rel.} {\bf 7}
  (2004) 5--56}, [\href{http://arxiv.org/abs/gr-qc/0311082}{{\tt
  gr-qc/0311082}}].

\bibitem{Burgess:2009ea}
C.~P. Burgess, H.~M. Lee and M.~Trott, \emph{{Power-counting and the Validity
  of the Classical Approximation During Inflation}},
  \href{http://dx.doi.org/10.1088/1126-6708/2009/09/103}{\emph{JHEP} {\bf 09}
  (2009) 103}, [\href{http://arxiv.org/abs/0902.4465}{{\tt 0902.4465}}].

\bibitem{Chluba:2015bqa}
J.~Chluba, J.~Hamann and S.~P. Patil, \emph{{Features and New Physical Scales
  in Primordial Observables: Theory and Observation}},
  \href{http://dx.doi.org/10.1142/S0218271815300232}{\emph{Int. J. Mod. Phys.}
  {\bf D24} (2015) 1530023}, [\href{http://arxiv.org/abs/1505.01834}{{\tt
  1505.01834}}].

\bibitem{Linde:1983gd}
A.~D. Linde, \emph{{Chaotic Inflation}},
  \href{http://dx.doi.org/10.1016/0370-2693(83)90837-7}{\emph{Phys. Lett.} {\bf
  B129} (1983) 177--181}.

\bibitem{Maeda:1987xf}
K.-i. Maeda, \emph{{Inflation as a Transient Attractor in R**2 Cosmology}},
  \href{http://dx.doi.org/10.1103/PhysRevD.37.858}{\emph{Phys. Rev.} {\bf D37}
  (1988) 858}.

\bibitem{Muller:1989rp}
V.~Muller, H.~J. Schmidt and A.~A. Starobinsky, \emph{{Power law inflation as
  an attractor solution for inhomogeneous cosmological models}},
  \href{http://dx.doi.org/10.1088/0264-9381/7/7/012}{\emph{Class. Quant. Grav.}
  {\bf 7} (1990) 1163--1168}.

\bibitem{GarciaBellido:1995kc}
J.~Garcia-Bellido and D.~Wands, \emph{{General relativity as an attractor in
  scalar - tensor stochastic inflation}},
  \href{http://dx.doi.org/10.1103/PhysRevD.52.5636}{\emph{Phys. Rev.} {\bf D52}
  (1995) 5636--5642}, [\href{http://arxiv.org/abs/gr-qc/9503049}{{\tt
  gr-qc/9503049}}].

\bibitem{Ferreira:1997hj}
P.~G. Ferreira and M.~Joyce, \emph{{Cosmology with a primordial scaling
  field}}, \href{http://dx.doi.org/10.1103/PhysRevD.58.023503}{\emph{Phys.
  Rev.} {\bf D58} (1998) 023503},
  [\href{http://arxiv.org/abs/astro-ph/9711102}{{\tt astro-ph/9711102}}].

\bibitem{DUFF1985355}
M.~Duff and C.~Pope, \emph{Consistent truncations in kaluza-klein theories},
  \href{http://dx.doi.org/http://dx.doi.org/10.1016/0550-3213(85)90140-3}{\emph{Nuclear
  Physics B} {\bf 255} (1985) 355 -- 364}.

\bibitem{Gibbons:2003gp}
G.~W. Gibbons and C.~N. Pope, \emph{{Consistent S**2 Pauli reduction of
  six-dimensional chiral gauged Einstein-Maxwell supergravity}},
  \href{http://dx.doi.org/10.1016/j.nuclphysb.2004.07.016}{\emph{Nucl. Phys.}
  {\bf B697} (2004) 225--242}, [\href{http://arxiv.org/abs/hep-th/0307052}{{\tt
  hep-th/0307052}}].

\bibitem{Weinberg:1972kfs}
S.~Weinberg, \emph{{Gravitation and Cosmology}}.
\newblock John Wiley and Sons, New York, 1972.

\bibitem{Burgess:2015lda}
C.~P. Burgess, R.~Diener and M.~Williams, \emph{{Self-Tuning at Large
  (Distances): 4D Description of Runaway Dilaton Capture}},
  \href{http://dx.doi.org/10.1007/JHEP10(2015)177}{\emph{JHEP} {\bf 10} (2015)
  177}, [\href{http://arxiv.org/abs/1509.04209}{{\tt 1509.04209}}].

\bibitem{Polchinski:1995sm}
J.~Polchinski and A.~Strominger, \emph{{New vacua for type II string theory}},
  \href{http://dx.doi.org/10.1016/S0370-2693(96)01219-1}{\emph{Phys. Lett.}
  {\bf B388} (1996) 736--742}, [\href{http://arxiv.org/abs/hep-th/9510227}{{\tt
  hep-th/9510227}}].

\bibitem{Aghababaie:2002be}
Y.~Aghababaie, C.~P. Burgess, S.~L. Parameswaran and F.~Quevedo, \emph{{SUSY
  breaking and moduli stabilization from fluxes in gauged 6-D supergravity}},
  \href{http://dx.doi.org/10.1088/1126-6708/2003/03/032}{\emph{JHEP} {\bf 03}
  (2003) 032}, [\href{http://arxiv.org/abs/hep-th/0212091}{{\tt
  hep-th/0212091}}].

\bibitem{Braun:2006se}
A.~P. Braun, A.~Hebecker and M.~Trapletti, \emph{{Flux Stabilization in 6
  Dimensions: D-terms and Loop Corrections}},
  \href{http://dx.doi.org/10.1088/1126-6708/2007/02/015}{\emph{JHEP} {\bf 02}
  (2007) 015}, [\href{http://arxiv.org/abs/hep-th/0611102}{{\tt
  hep-th/0611102}}].

\bibitem{Adam:2015rua}
{\scshape Planck} collaboration, R.~Adam et~al., \emph{{Planck 2015 results. I.
  Overview of products and scientific results}},
  \href{http://arxiv.org/abs/1502.01582}{{\tt 1502.01582}}.

\bibitem{Albrecht:2001xt}
A.~Albrecht, C.~P. Burgess, F.~Ravndal and C.~Skordis, \emph{{Natural
  quintessence and large extra dimensions}},
  \href{http://dx.doi.org/10.1103/PhysRevD.65.123507}{\emph{Phys. Rev.} {\bf
  D65} (2002) 123507}, [\href{http://arxiv.org/abs/astro-ph/0107573}{{\tt
  astro-ph/0107573}}].

\bibitem{Conlon:2008cj}
J.~P. Conlon, R.~Kallosh, A.~D. Linde and F.~Quevedo, \emph{{Volume Modulus
  Inflation and the Gravitino Mass Problem}},
  \href{http://dx.doi.org/10.1088/1475-7516/2008/09/011}{\emph{JCAP} {\bf 0809}
  (2008) 011}, [\href{http://arxiv.org/abs/0806.0809}{{\tt 0806.0809}}].

\bibitem{Kofman:2004yc}
L.~Kofman, A.~D. Linde, X.~Liu, A.~Maloney, L.~McAllister and E.~Silverstein,
  \emph{{Beauty is attractive: Moduli trapping at enhanced symmetry points}},
  \href{http://dx.doi.org/10.1088/1126-6708/2004/05/030}{\emph{JHEP} {\bf 05}
  (2004) 030}, [\href{http://arxiv.org/abs/hep-th/0403001}{{\tt
  hep-th/0403001}}].

\end{thebibliography}\endgroup

\end{document}